\documentclass[preprint,twocolumn]{IEEEtran}
\usepackage{amsmath,amsfonts}
\usepackage{algorithmic}
\usepackage{algorithm}
\usepackage{array}
\usepackage[caption=false,font=normalsize,labelfont=sf,textfont=sf]{subfig}
\usepackage{textcomp}
\usepackage{stfloats}
\usepackage{url}
\usepackage{verbatim}
\usepackage{graphicx}
\usepackage{cite}
\usepackage{eso-pic}
\usepackage{epsfig}
\usepackage{bm}

\usepackage{amssymb}

\begin{document}

\title{Edge turbulence controlled by topologically self-optimized fluxes in fusion devices}

\author{A. Bershadskii\textsuperscript{1}, 

\thanks{\textsuperscript{1}ICAR, P.O. Box 31155, Jerusalem 91000, Israel}%
}

\maketitle

\begin{abstract}

The magnetic field topology undergoes a drastic transition near the separatrix in magnetic confinement fusion devices, inducing topological chaos. This phenomenon is investigated across tokamaks, stellarators, and the RFX-mod reversed-field pinch. Within this boundary region, self-optimized fluxes of shear-generated magnetic and cross-helicity, together with the ensemble-averaged magnetic moment, are hypothesized to govern plasma stability, self-organization, and transport. By treating the averaged magnetic moment as an active scalar and using the Fokker–Planck equation to reconcile kinetic and fluid descriptions, this study couples its dynamics with a novel multichannel, self-optimized helicity cascade loop. The derived spectral laws for the floating potential and ion saturation current agree with measurements from numerous experiments. Distributed chaos theory quantifies the randomness of the observed states, revealing that a narrow layer at the separatrix suppresses strong core randomness, transforming it into distributed chaos in the scrape-off layer. A constrained variational principle for active control has also been considered. Tuning Lagrangian multipliers via helicity injection drives the system toward an attractor state that stabilizes the separatrix shear layer, suppresses edge-localized modes, and spreads the heat load, using magnetic reconnection as a nonlinear solver. This framework circumvents conventional diffusive limitations, providing an analytical foundation to model heat-load patterns and stochastic ripple loss in next-step devices.

\end{abstract}

\section{Introduction}  

   Currently, a significant issue hindering the practical use of magnetic confinement fusion devices is the chaotic/turbulent processes in their edge dynamics—the region where closed magnetic field lines transition to open ones, generating topological chaos (see, for instance, Refs. \cite{ksk,boozer} and references therein). These processes result in anomalous particle and energy loss to the reactor walls. Although the most accessible measurements are conducted in this domain, there is no practically feasible theory or comprehensive phenomenology for such processes that can be successfully compared with the existing measurements. \\

  The power spectra of observable dynamical characteristics (e.g., ion saturation current 
 and floating potential) at the edge of magnetically confined fusion devices, such as tokamaks, stellarators, and reversed-field pinches, exhibit an intricate and poorly understood variability \cite{ped}. In these experimental devices, the number of factors affecting the dynamics is large. Therefore, determining in advance which parameters are significant when analyzing the spectra of edge turbulent fluctuations across various devices is challenging, particularly due to the lack of a robust theory of turbulent dynamics. \\

This problem stimulates the development of 3D fluid simulations of edge turbulence (see for recent reviews Refs. \cite{gr,schw}), which include recent models incorporating electromagnetic effects. Certain phenomenological insights can serve as an effective connection between numerical simulations and a comprehensive theory. In particular, the intersection of steep pressure gradients, magnetic ripple, and Resonant Magnetic Perturbations (RMPs) creates a region of intrinsic stochasticity near the separatrix \cite{gao}. In these topologically chaotic layers (Pedestal and near-SOL), the magnetic moment $\mu$ (which can be treated as an adiabatic invariant in the Core domain of the fusion devices) undergoes rapid, phase-dependent fluctuations that render classical Hamiltonian conservation laws and simple diffusive models insufficient \cite{via1}. The consideration of the flux of averaged magnetic moment $\varepsilon_{\mu}$, as a statistical quasi-invariant governing a cascade process in the phase space, along with a multichannel, self-optimized, cascade loop of the helicities spontaneously (or intentionally) generated by a combination strong shear layer and magnetic topology near the separatrix, enables a link between kinetic and fluid dynamic descriptions. \\

    The notion of smoothness can help in the quantitative grading of the chaotic/turbulent regimes based on their level of randomness. The spectral analysis can be utilized for this objective. Specifically, stretched exponential spectra 
\begin{equation}
E(f) \propto \exp-(f/f_{\beta})^{\beta}
\end{equation}   
 are anticipated for the smooth magnetohydrodynamics \cite{ber2} (here $1 \geq \beta > 0$ and $f$ is the frequency), and 
pure exponential spectra 
\begin{equation} 
E(f) \propto \exp(-f/f_c),  
\end{equation}  
are anticipated for deterministic chaos \cite{oh,mm1,mm2,mm3,kds}. \\

  For $1 > \beta$, the behavior remains smooth yet is not deterministic (this will be referred to as distributed chaos; further explanation of the term follows). It may also be regarded as soft turbulence \cite{wu}.
  The non-smooth dynamics (hard turbulence \cite{wu}) is usually defined by the power-law (scaling) spectra. \\
  
 In this method, the value of $\beta$ may be viewed as an appropriate indicator of the level of randomness. Specifically, the farther the parameter $\beta$ is from the deterministic $\beta = 1$ (i.e., the lower the $\beta$), the greater the level of randomness. \\ 

%%%%%%%%%%%%%%% 1 %%%%%%%%%%%%%%%%%%
\begin{figure} \vspace{-0.4cm}\centering
\epsfig{width=.47\textwidth,file=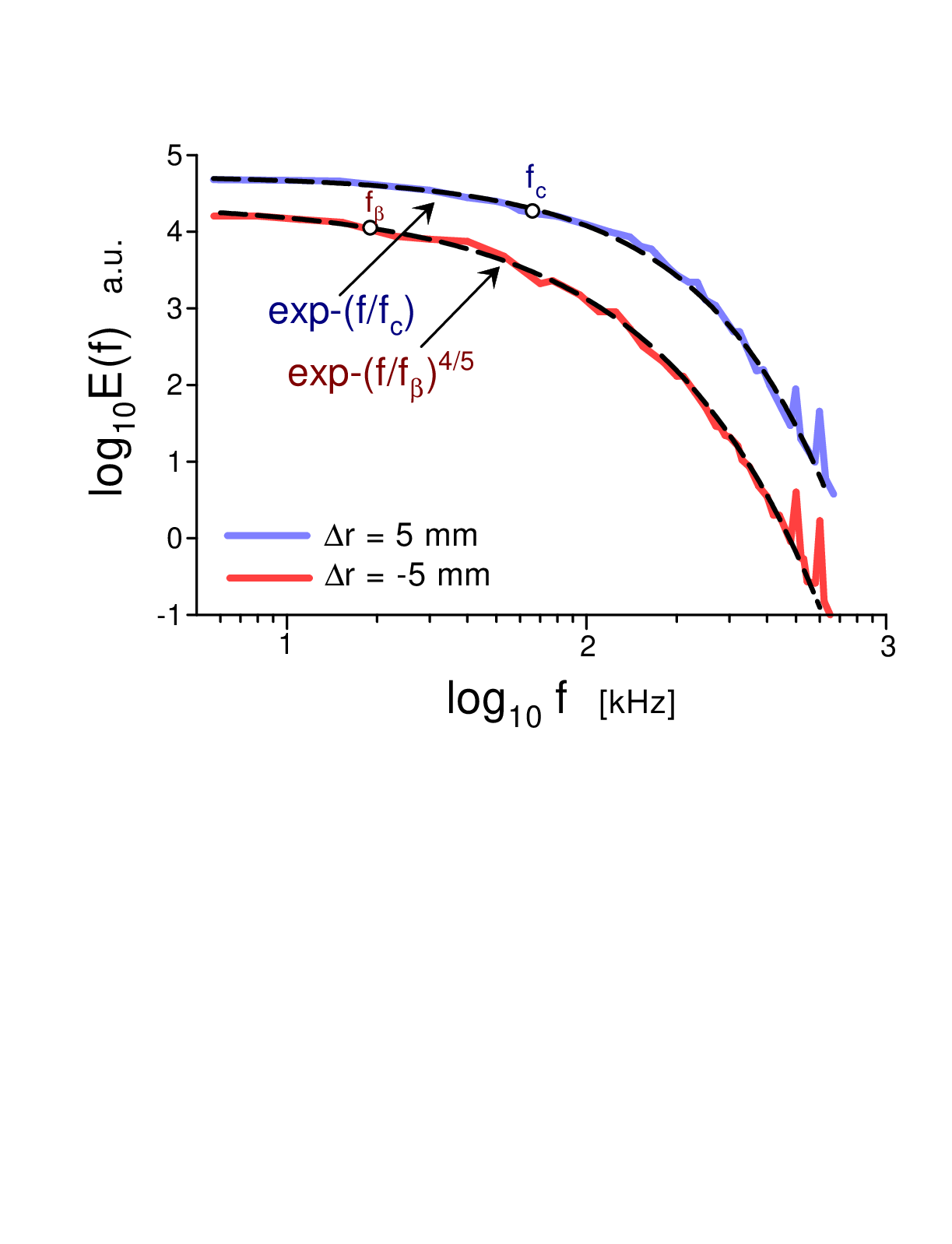} \vspace{-4.8cm}
\caption{Power spectra of the floating potential time series measured at two different radial positions (with respect to the separatrix location at $r_{sep}$): $\Delta r = \pm 5$ mm, where $\Delta r = r-r_{sep}$, for the ISTTOK tokamak, The spectra are vertically shifted for clarity.} 
\end{figure}
%%%%%%%%%%%%%%%%%%%%%%%%%%%%%%%%%%%   

 Let us consider some examples. The first example is taken from measurements of the floating potential time variability carried out in the edge plasma using a Langmuir probe (as in most examples provided in the paper). In this case, a small-scale circular cross-section tokamak, ISTTOK, was exploited in the H-plasma regime \cite{dud} (note that in the H-mode, electromagnetic fluctuations, such as Drift-Alfv\'en waves and Kinetic Ballooning Modes, become critical).
 
 Figure 1 shows the power spectra of the floating potential time series measured at two different radial positions (with respect to the separatrix location at $r_{sep}$): $\Delta r = \pm 5$ mm, where $\Delta r = r-r_{sep}$. The spectral data were taken from Fig. 10 of the Ref. \cite{dud}.
   The dashed curves in Fig. 1 indicate the best fits corresponding to the stretched exponential Eq. (1) for $\Delta r = -5$ mm  ($\beta = 4/5 < 1$, distributed chaos), and to the exponential Eq. (2) ($\beta =1$, deterministic chaos) for  $\Delta r =5$ mm.\\

%%%%%%%%%%%%%%% 2 %%%%%%%%%%%%%%%%%%
\begin{figure} \vspace{+0.5cm}\centering
\epsfig{width=.46\textwidth,file=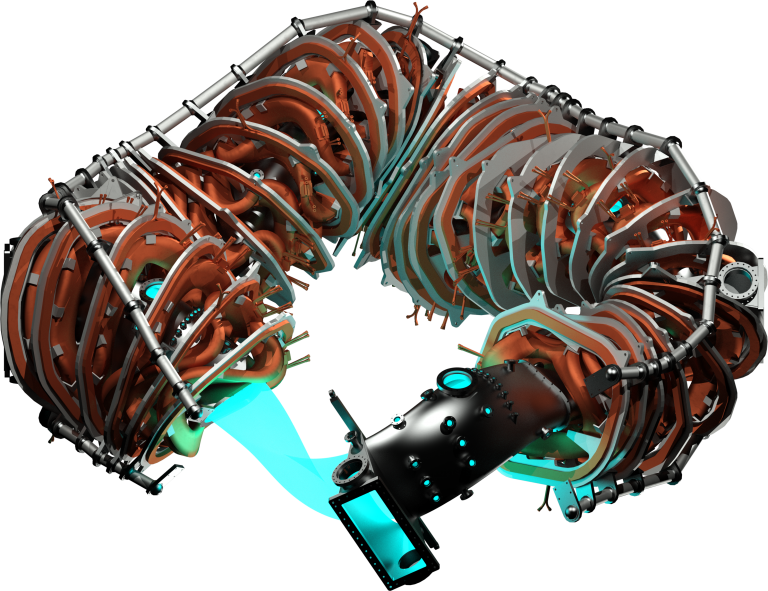} \vspace{-0cm}
\caption{ A sketch of the HSX stellarator.}
\end{figure}
%%%%%%%%%%%%%%%%%%%%%%%%%%%%%%%%%%% 

  The second example comes from the floating potential time series measured at various radial locations inside the edge region of the HSX stellarator, which exhibits quasi-helical symmetry. Figure 2 (adapted from the site \cite{HSX}) shows a sketch of the HSX stellarator. The stellarator uses 48 modular coils to generate a dominant, single helical magnetic component, reducing particle loss and enabling studies of plasma turbulence and transport in a regime similar to tokamaks but without their inherent instabilities. Its magnetic field has been optimized so that its strength is approximately symmetric in a helical direction, which makes its magnetic confinement similar to that of an axisymmetric tokamak. This type of symmetry is supposed to reduce particle and energy transport in the stellarator. Moreover, due to the neoclassical optimization of the HSX stellarator, the turbulence becomes the dominating source of transport (see, for instance, Ref. \cite{ger} and references therein). The 3D-shaped modular coils create the optimized magnetic field, allowing for studies of areas critical to fusion device design, like plasma-wall interactions and turbulence. The HSX stellarator confines plasma hotter than 10 million levels Celsius (heated by Electron Cyclotron Resonance Heating) using a 1 Tesla magnetic field.\\

%%%%%%%%%%%%%%% 3 %%%%%%%%%%%%%%%%%%
\begin{figure} \vspace{-0.5cm}\centering 
\epsfig{width=.45\textwidth,file=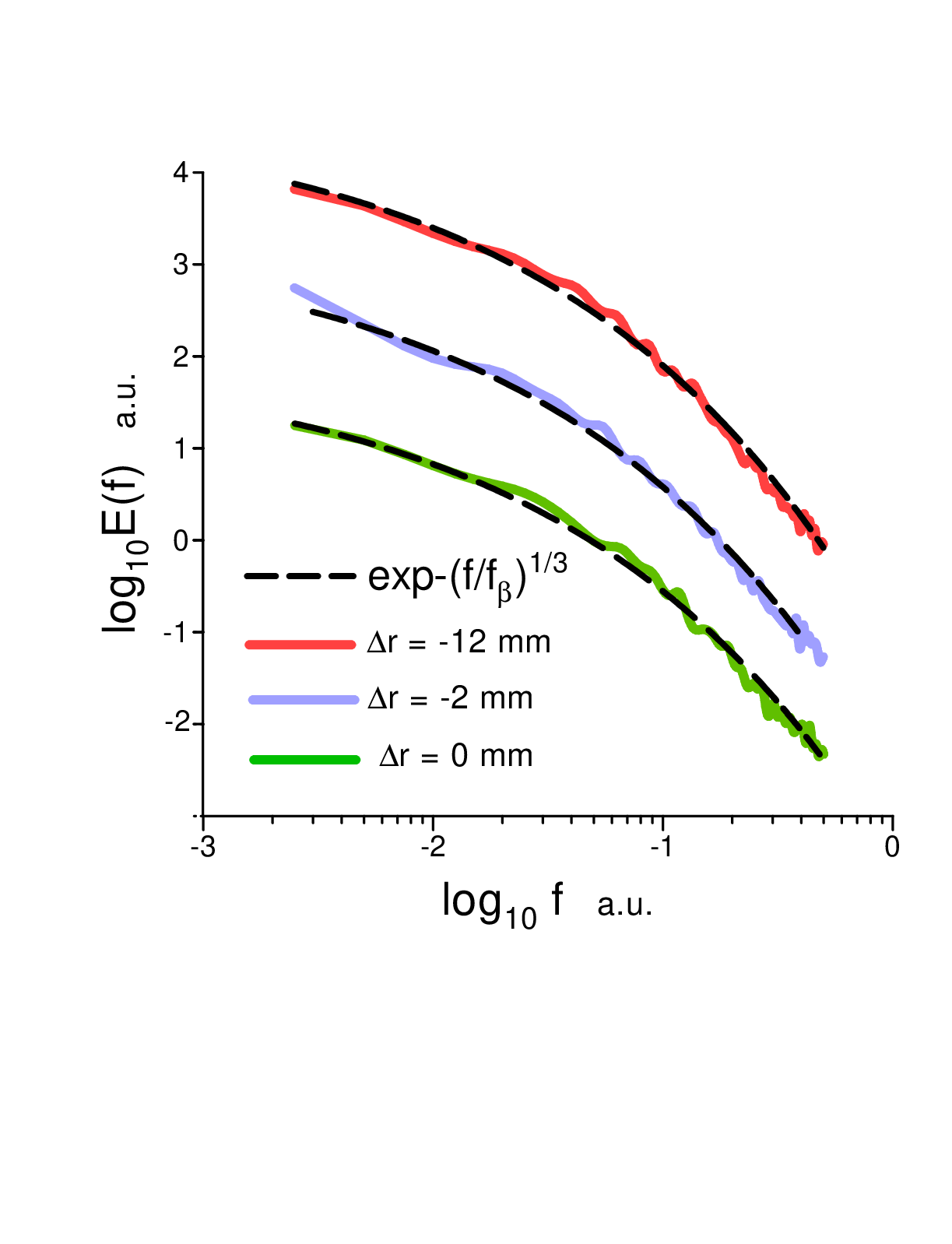} \vspace{-2.9cm}
\caption{Power spectra of the floating potential time series measured at different radial positions (with respect to the separatrix location at $r_{sep}$): $\Delta r = r-r_{sep}$, for HSX stellarator.  The spectra are vertically shifted for clarity.} 
\end{figure}
%%%%%%%%%%%%%%%%%%%%%%%%%%%%%%%%%%% 
   
  Figures 3 and 4 show the power spectra of the floating potential time series measured at various radial positions (with respect to the separatrix location at $r_{sep}$): $\Delta r = r-r_{sep}$. The time series were taken from the site of International Stellarator/Heliotron Profile DataBase (ISHPDB)\footnote{\url{https://ishpdb.ipp-hgw.mpg.de/ISHPDB\_public/physicsTopics/edge\_turbulence/index.html}}. The spectra were calculated utilizing the maximum entropy method, yielding optimal resolution for chaotic time series \cite{oh}. In Fig. 3, the spectra are shown for negative $\Delta r$ (inside separatrix), whereas in Fig. 4  the spectra are shown for positive $\Delta r$ (outside separatrix). \\
  
  One can see that, similar to the first example, randomness is greater ($\beta$ is smaller) inside the separatrix than outside it. This information is already interesting on its own. But funding the reasons for this phenomenon will provide a deeper understanding of randomization in the chaotic/turbulent edge plasma. In the subsequent sections, the observed values of $\beta$ will be related to the topological quasi-invariant fluxes of non-barotropic (and compressible) magnetohydrodynamics.\\

  When plasma pressure becomes so high that the magnetic field can no longer hold it rigidly (magnetic fluttering), the turbulence shifts from purely electrostatic to electromagnetic \cite{eic}. This transition occurs quite often when pushing for high-performance H-mode, a primary research objective. We should function close to this transition point since that is where the highest fusion power is produced, but we need to "ride the edge" without descending into complete instability. In the fusion devices, this can be understood literally because just the edge area of the devices is most sensitive to the transition. 
 
  %%%%%%%%%%%%%%% 4 %%%%%%%%%%%%%%%%%%
\begin{figure} \vspace{-0.7cm}\centering
\epsfig{width=0.46\textwidth,file=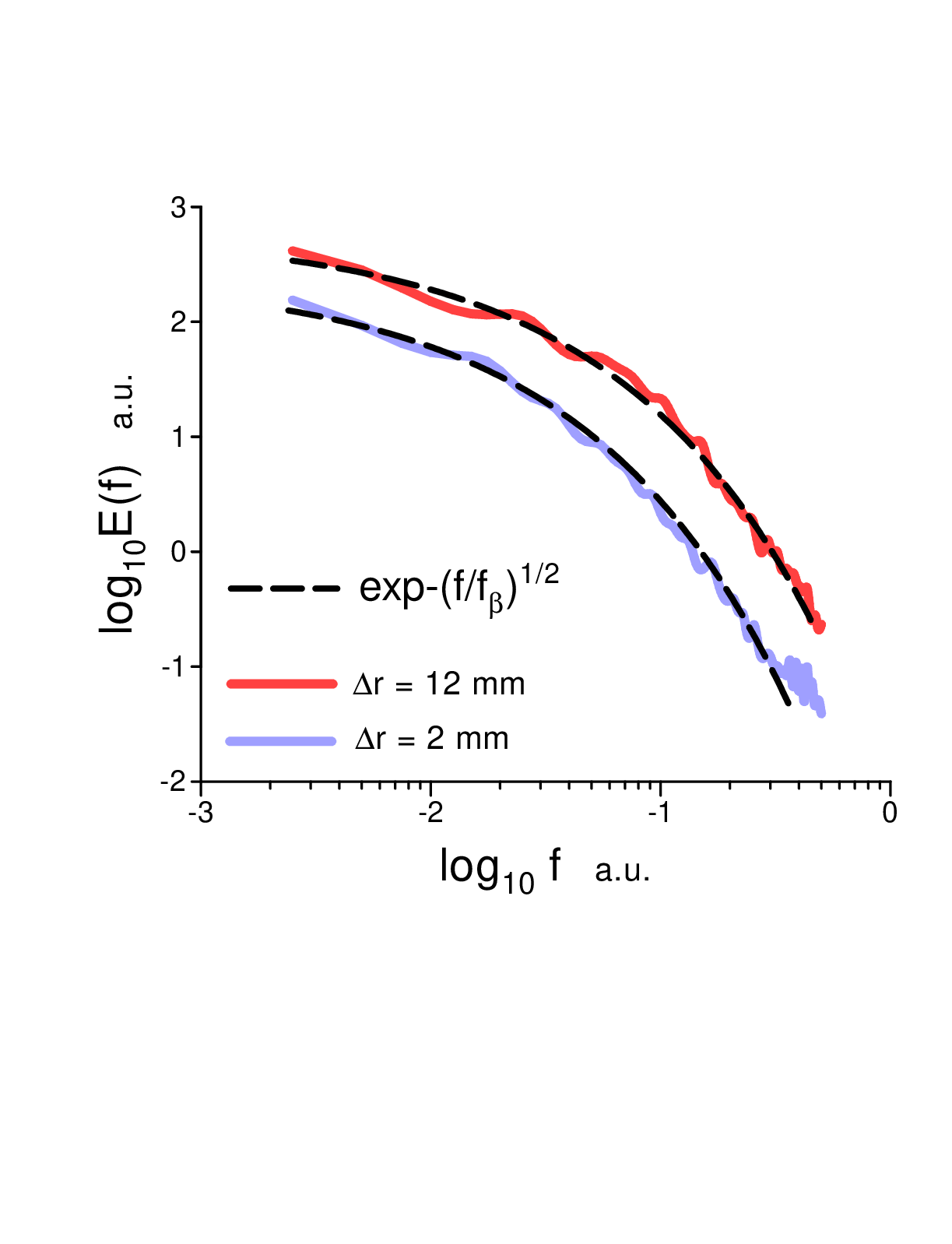} \vspace{-3.2cm}
\caption{As in Fig. 3 but for positive $\Delta r$.} 
\end{figure}
%%%%%%%%%%%%%%%%%%%%%%%%%%%%%%%%%%%      

 Approaching the separatrix significantly raises both the probability and strength of a shift from electrostatic to electromagnetic turbulence. 
 
 The local ratio of plasma pressure to magnetic pressure is very high relative to the "stiffness" of the magnetic field at the edge; the plasma possesses sufficient leverage to curve the magnetic field lines. This is the main birthplace of micro-tearing modes, which are entirely electromagnetic.
 
  In the core, magnetic field lines are closely coiled and structured. As we approach the separatrix, the magnetic shear (the level to which the field's twist changes as we move outward) tends to infinity. The high shear can suppress small electrostatic eddies, but it creates a perfect environment for electromagnetic instabilities \cite{con}. 
  
  The blobs (filaments)  begin as fluctuations of an electrostatic nature. As they approach the separatrix, they possess such local pressure that they expand ("balloon") outward, pulling the magnetic field lines along with them (the Electromagnetic Ballooning Mode) \cite{wie}.
  
  The intense pressure gradient in the Pedestal strip supplies energy for the electromagnetic instabilities, while magnetic reconnections (associated with the X-point, which lowers the energy barrier for magnetic reconnection) and Edge Localized Modes complete the picture. \\

   Finally, it should be noted that the transition from electrostatic to electromagnetic turbulence significantly influences the outcomes of measurements taken by the Langmuir probe on both sides of the separatrix. These effects are so pronounced  (especially the presence of magnetic "flutter" and fluctuating parallel currents) that these measurements often reflect not the turbulent characteristics of the plasma potential and plasma density (as would be anticipated in purely electrostatic turbulence, with the magnetic field as a fixed background), but an electrostatic response to an electromagnetic driver. In other words, the measured floating potential and ion saturation current can be governed by emerging electromagnetic dynamics (dominated by the topological fluxes) near the separatrix. In this paper, we will use the term ``edge'' to describe a certain layer centered at the separatrix. This is also a shear layer generating the kinetic helicity $H_k$ that the blobs then carry away into SOL and the layer where flux of the ensemble-averaged magnetic moment (together with the fluxes related to the kinetic helicity) plays a significant role in the ion saturation current $I_{sat}$ measured by the Langmuir probe.  Actually, the ion saturation current $I_{sat}$ is locked to the (ion) kinetic layer. In contrast, the floating potential $\phi$ near the separatrix, which is almost entirely governed by the parallel mobility of much faster electrons, is locked to the edge magnetic layer and consequently is controlled by the quasi-invariant fluxes of magnetic and cross helicities. We will explore these phenomena in more detail below in the paper.

\section{MHD invariants and magneto-inertial range of scales}  
 
\subsection{MHD invariants}

  The ideal (non-dissipative) magnetohydrodynamics has three quadratic (fundamental) invariants: total energy, cross and magnetic helicity \cite{mt}. It is well known that these quantities play a significant role in magnetized plasma dynamics. 
 
   Magnetic helicity of a domain with spatial volume $V$ is
\begin{equation}
H_m = \int_{V} ({\bf A} ({\bf x},t) \cdot  {\bf B} ({\bf x},t)) ~ d{\bf x }
\end{equation}   
 where the magnetic field ${\bf B} = [{\nabla \times \bf A}]$ and $ {\bf A}$ is the magnetic potential ($\nabla \cdot {\bf A} =0$). Under certain boundary conditions, $H_m$ is an ideal invariant of magnetohydrodynamics \cite{mt}. There are some useful generalizations of magnetic helicity (see, for instance, \cite{shebalin} and references therein). The notion of relative magnetic helicity is especially useful for fusion devices such as tokamaks, stellarators, and reversed-field pinches \cite{mac} (further in the text, the term “relative” will be omitted, but will be implied).
 
 The fact that in dissipative plasma magnetic helicity generally decays more slowly than energy enables considering magnetic helicity as an adiabatic invariant in many practically important cases. \\ 
 
   The cross helicity of a domain with spatial volume $V$ is
\begin{equation}
H_{cr} = \int_{V} ({\bf u} ({\bf x},t) \cdot  {\bf B} ({\bf x},t)) ~ d{\bf x }
\end{equation}   
 
 The cross helicity is an invariant of the ideal magnetohydrodynamics. It has been shown in a paper \cite{ber1} that the second moment of the cross helicity $\rm{I}_{cr}$ is also an invariant of ideal magnetohydrodynamics (see also a review, Ref. \cite{sch}, where this statement has been broadened to include magnetic helicity). \\

   In ideal non-barotropic compressible magnetohydrodynamics, the magnetic helicity is conserved. In contrast, the standard MHD cross helicity is not conserved because the entropy and temperature gradients are not aligned (\(\nabla S\times \nabla T\ne 0\)), producing sink or source terms To resolve that issue, one can use a replacement ${\bf u} \rightarrow {\bf u}_{t}={\bf u}-\sigma \nabla S$ (where  $d\sigma /dt=T$)\cite{y},\cite{sy}. \\
   
   In fusion devices like tokamaks and stellarators, the magnetic and cross helicities are spontaneously generated in the plasma shear layers around the separatrix due to large-scale anisotropy (driven by the strong toroidal magnetic field), micro-instabilities, and the interaction of the strong gradients.

 \subsection{Magneto-inertial range of scales}
 
 It was suggested in the Ref. \cite{ber2} to use the MHD invariants and a Kolmogorov-like phenomenology to relate the characteristic value of the magnetic field $B_c$ to the characteristic frequency $f_c$ (the concept of magneto-inertial range of scales). In the present paper, this approach will be applied (adapted) to the characteristic values of the floating potential $\phi_c$ and ion saturation current $I_{sat,c}$.\\
 
  Namely, for the magneto-inertial range of scales dominated by cross helicity and Kolmogov's parameter $\varepsilon$ (the constant total energy flux), the Kolmogorov-like dimensional considerations (adapted to the characteristic floating potential $\phi_c$ instead of $B_c$) result in a scaling relationship
\begin{equation}
\phi_c \propto \left|\frac{d H_{cr}}{dt}\right| \varepsilon^{-1}f_c^{2}   
\end{equation}

     For the magneto-inertial range of scales dominated by magnetic helicity and Kolmogov's parameter $\varepsilon$ analogous dimensional considerations result in a scaling relationship
 \begin{equation}
\phi_c \propto \left|\frac{d H_m}{dt}\right|^{1/2} \varepsilon^{0}f_c^{1/2}   
\end{equation}

    For the magneto-inertial range of scales dominated by the flux of the second moment of cross helicity $d{\rm I}_{cr}/dt$  and Kolmogov's parameter $\varepsilon$ analogous dimensional considerations result in a scaling relationship

\begin{equation}
\phi_c \propto \left|\frac{d {\rm I}_{cr}}{dt}\right|^{1/2} \varepsilon^{-1/4}f_c^{1/4}   
\end{equation}  

\section{Distributed chaos in fluctuations of the floating potential}
 
  A strong alteration in the deterministically chaotic condition of a system may result in the random variability of the characteristic scale $k_c$ in the exponential spectrum Eq. (2) which can subsequently lead to a transition toward a non-deterministic state of the system (see, for instance, Fig. 1). In this situation, it is necessary to apply an ensemble-averaging approach to calculate the power spectra:
\begin{equation}
E(f) \propto \int_0^{\infty} P(f_c) \exp -(f/f_c)df_c 
\end{equation}    
with a probability distribution $P(f_c)$ representing the random fluctuations of $f_c$. Therefore, the emerging smooth, non-deterministic chaotic dynamics will be referred to as `distributed chaos'.\\ 

   Let us first determine $P(f_c)$ for the magneto-inertial range of scales, which is dominated by magnetic helicity. If the characteristic floating potential $\phi_c$ has the half-normal probability distribution $P(\phi_c) \propto \exp- (\phi_c^2/2\sigma^2)$ (i.e., a zero mean normal distribution truncated to have nonzero probability density for positive values of its argument only: if $\phi$ has a normal distribution,  then $\phi = |\phi|$ has a half-normal one \cite{jkb}), then it follows from Eq. (6) that $f_c$ has the chi-squared ($\chi^{2}$) probability distribution
\begin{equation}
P(f_c) \propto f_c^{-1/2} \exp-(f_c/4f_{\beta})  
\end{equation}  
where $f_{\beta}$ is a constant. This probability distribution for the characteristic value of the floating potential (and ion saturated current) should not be confused with the probability distribution of the measured values of the fluctuating floating potential $\phi (t)$, which can be non-normal. The situation is well known for the statistical properties of the turbulent variables \cite{ber2,my}.

   Substitution of Eq. (9) into Eq. (8) results in an equation
\begin{equation}
E(f) \propto \exp-(f/f_{\beta})^{1/2}  
\end{equation}  
  
  We have already seen the spectrum Eq. (10) in Fig. 4 for the HSX stellarator outside the separatrix.  Because the plasma dynamics of the HSX stellarator is characterized by a uniquely configured helical magnetic symmetry to reduce neoclassical transport, the inherent magnetic helicity governs the "self-organization" of this turbulence into large-scale structures (like blobs and magnetic islands) that carry energy and particles across the separatrix. This may explain the emergence of the spectrum Eq. (10) (cf. also for the appearance of this spectrum for magnetic energy in Refs.  \cite{ber2,ber3}).\\

   An analogous power spectrum was observed for the ASDEX Upgrade tokamak with a specially improved Langmuir probe. The measurements were produced a few millimeters outside the separatrix ($\Delta r > 0$ cf Fig. 4). Figure 5 shows the power spectrum of the floating potential fluctuations (the spectral data were taken from Fig. 14c of the Ref. \cite{nold}).  The dashed curve (the best fit) indicates correspondence to Eq. (10). The vertical dotted arrow indicates the position of $f_\beta$. The magneto-inertial range begins around 10 kHz.

 \section{Magneto-inertial range of scales dominated by cross helicity}      
 
   Let us look for the spectrum of the distributed chaos as a stretched exponential (see Introduction and Eq. (8))
\begin{equation}
E(f) \propto \int_0^{\infty} P(f_c) \exp -(f/f_c)dk_c \propto \exp-(f/f_{\beta})^{\beta} 
\end{equation}  

  For large $f_c$ the probability distribution $P(f_c) $ can be estimated using Eq. (11)  \cite{jon}
\begin{equation}
P(f_c) \propto f_c^{-1 + \beta/[2(1-\beta)]}~\exp(-\gamma f_c^{\beta/(1-\beta)}) 
\end{equation}  
  
   On the other hand, the estimates Eqs. (5-7) can be generalized
\begin{equation}
 \phi_c \propto f_c^{\alpha}   
\end{equation}  
   For the half-normally distributed $\phi_c$ a relationship between $\beta$ and $\alpha$ can be derived from equations (12) and (13)
\begin{equation}
\beta = \frac{2\alpha}{1+2\alpha}  
\end{equation}  
   
   Since for the magneto-inertial range, dominated by the cross helicity, $\alpha =2$ (see Eq. (5)), the power spectrum of the floating potential can be estimated as
\begin{equation}
 E(f) \propto \exp-(f/f_{\beta})^{4/5}  
\end{equation}  

   The spectrum Eq. (15) can be recognized (inside the separatrix) in Fig. 1. In devices like ISTTOK, poloidal asymmetries in floating potential fluctuations suggest that the cross helicity in the edge region can dominate the distributed chaos/turbulence and particle flux (see Introduction).\\

  %%%%%%%%%%%%%%% 5 %%%%%%%%%%%%%%%%%%
\begin{figure} \vspace{-0.7cm}\centering
\epsfig{width=.42\textwidth,file=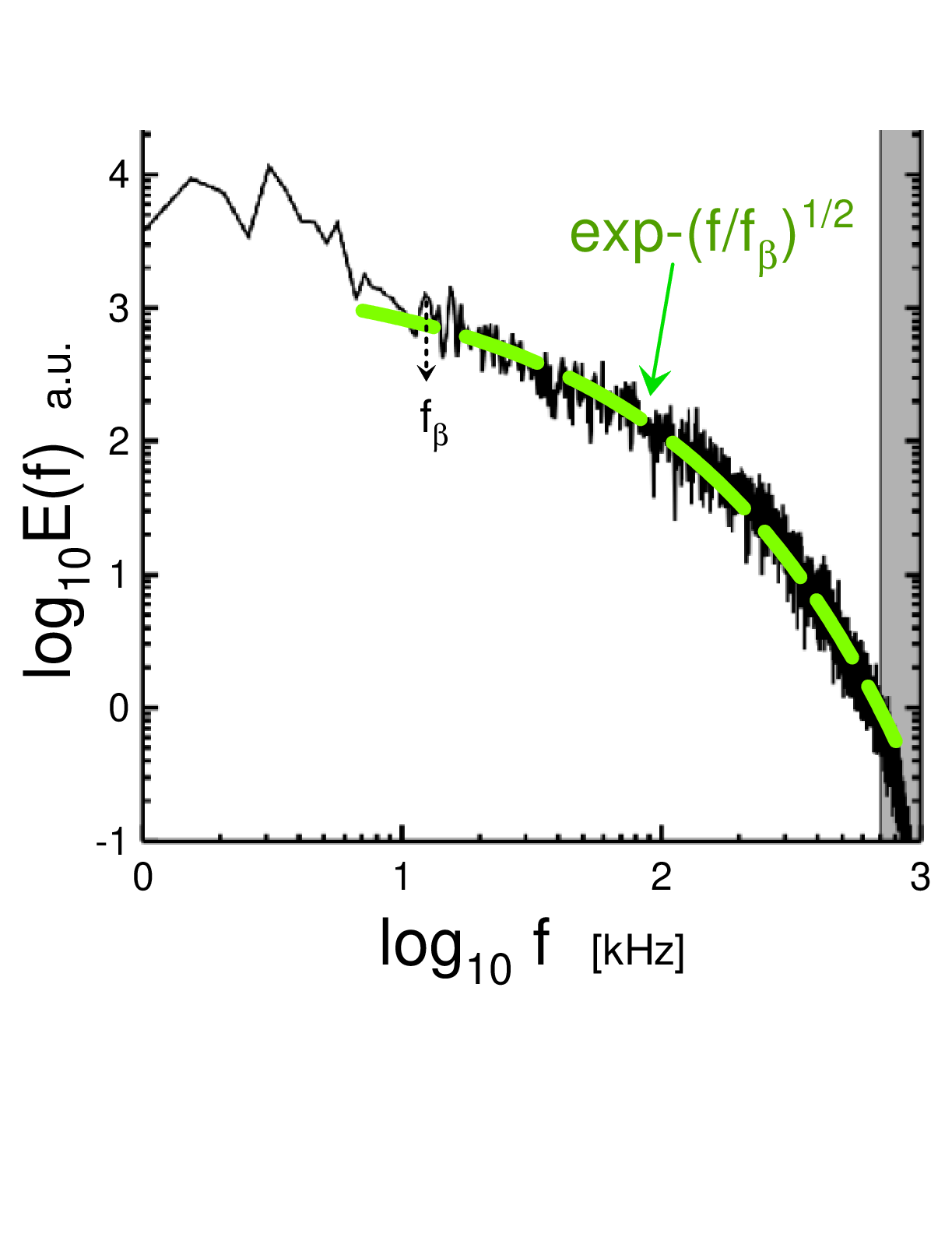} \vspace{-2.3cm}
\caption{Power spectrum of the floating potential fluctuations for ASDEX Upgrade tokamak. The measurements were produced a few millimeters outside the separatrix ($\Delta r > 0$). The cut-off frequency of the anti-aliasing filters are shown by the vertical gray band.\\} 
\end{figure}
%%%%%%%%%%%%%%%%%%%%%%%%%%%%%%%%%%% 
 %%%%%%%%%%%%%%% 6 %%%%%%%%%%%%%%%%%%
\begin{figure} \vspace{-0.4cm}\centering
\epsfig{width=.49\textwidth,file=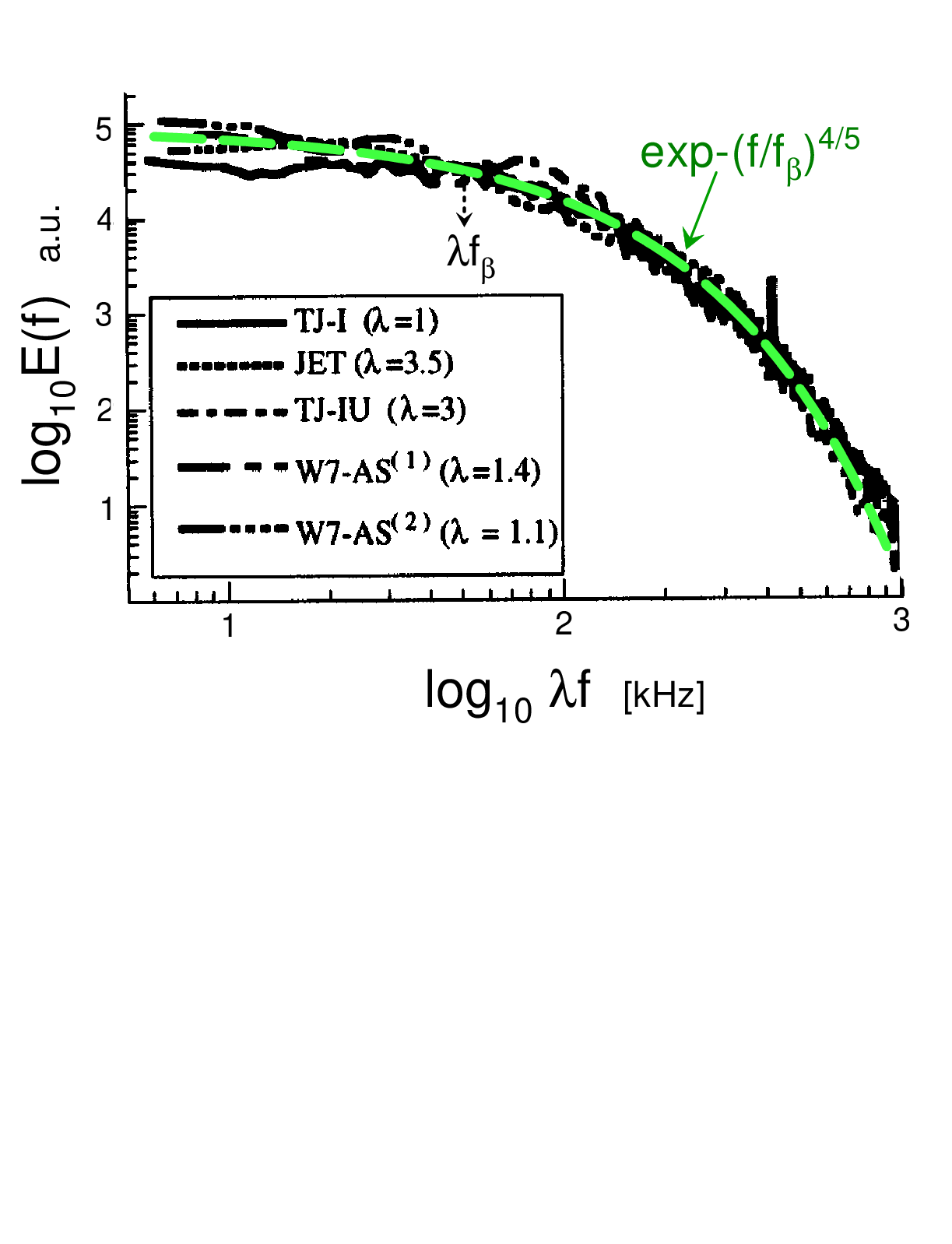} \vspace{-5.4cm}
\caption{Rescaled power spectra of fluctuations of the floating potential for different tokamak and stellarator types.} 
\end{figure}
%%%%%%%%%%%%%%%%%%%%%%%%%%%%%%%%%%% 

  The examples above show that the power spectra of floating-potential fluctuations can have different values of $\beta \leq 1$, depending on the specific device and the probe's position relative to the separatrix. In a seminal paper \cite{ped}, however, the spectra's universal behavior was empirically explored in the edge turbulence in the form
\begin{equation}
E(f) = E_0 g(\lambda f)   
\end{equation}  
where the parameter $\lambda$ is determined solely by the device type (regardless of the probe's position relative to the separatrix). 

  Figure 6 shows the results of this attempt for some tokamak and stellarator configurations. The spectral data were taken from the Ref. \cite{ped}. The dashed curve in this figure shows a stretched exponential Eq. (15) for reference.

\section{Magneto-inertial  range dominated by second moment of cross helicity}   
   
    Since for the magneto-inertial range, dominated by the second moment of cross helicity, $\alpha =1/4$ (see Eq. (7)), the power spectrum of the floating potential can be estimated (using Eq. (14)) as
\begin{equation}
 E(f) \propto \exp-(f/f_{\beta})^{1/3}  
\end{equation}  
  
   The spectrum Eq. (17) can be recognized for HSX stellarator inside (and at) the separatrix Fig. 3 (see also Introduction). \\ 
   
  The most practical way to measure magnetic fields in a high-radiation, high-temperature plasma environment is to use a simple loop of wire (inductive pick-up coils). According to Faraday’s Law of induction, a changing magnetic field through a coil induces a `Voltage' proportional to the time derivative of that field. Therefore, it is also interesting to estimate the power spectrum of the time derivative of the magnetic field in the magneto-inertial range, dominated by the second moment of cross helicity, as it can be directly compared with the floating potential.

 %%%%%%%%%%%%%%% 7 %%%%%%%%%%%%%%%%%%
\begin{figure} \vspace{-0.8cm}\centering \hspace{-1cm}
\epsfig{width=.47\textwidth,file=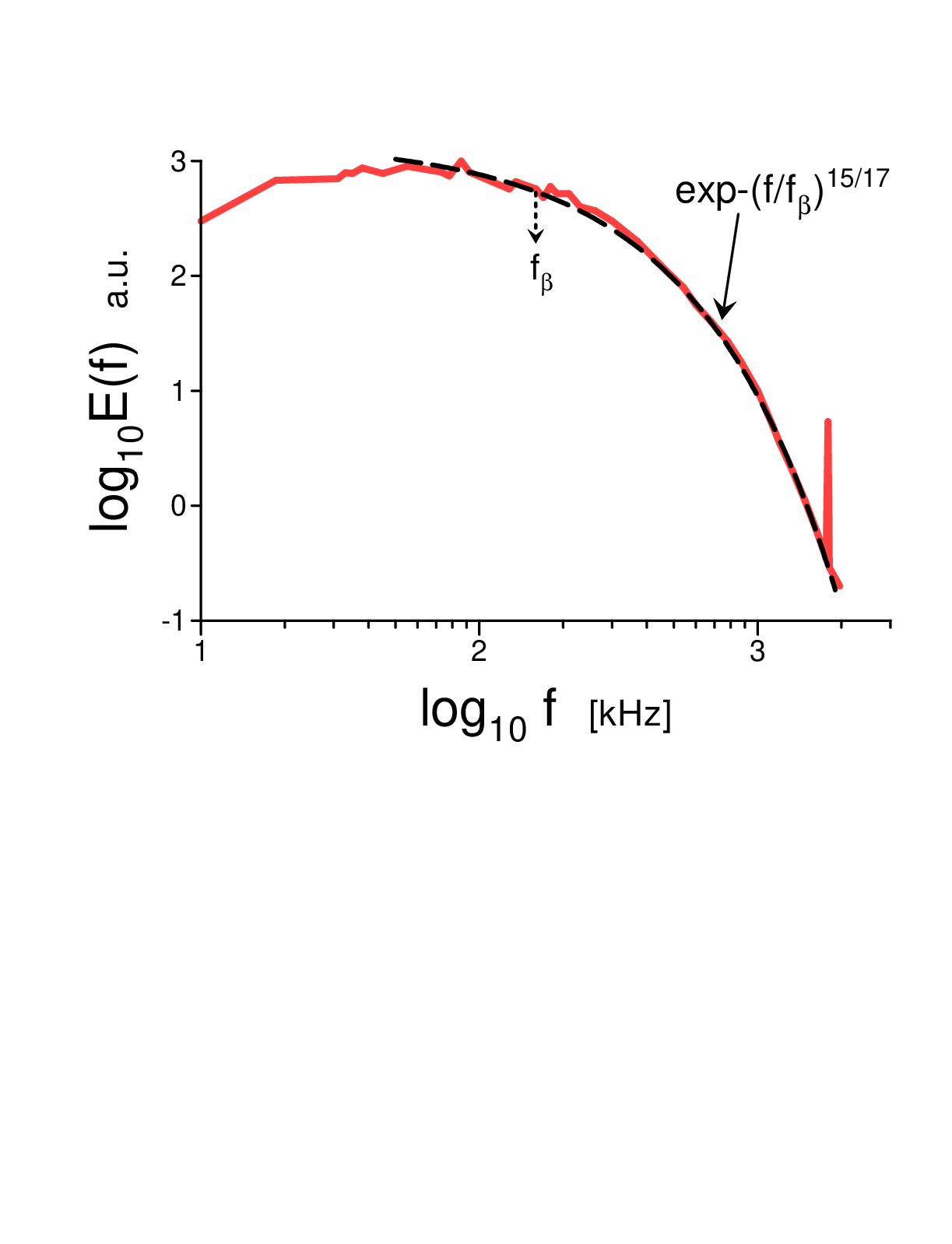} \vspace{-4.7cm}
\caption{Power spectrum of the time derivative of the toroidal magnetic field measured in the chaotic plasma edge of the RFX-mod reversed field pinch in the Quasi Single Helical state.} 
\end{figure}
%%%%%%%%%%%%%%%%%%%%%%%%%%%%%%%%%%%  
%%%%%%%%%%%%%%% 8 %%%%%%%%%%%%%%%%%%
\begin{figure} \vspace{-0cm}\centering
\epsfig{width=.45\textwidth,file=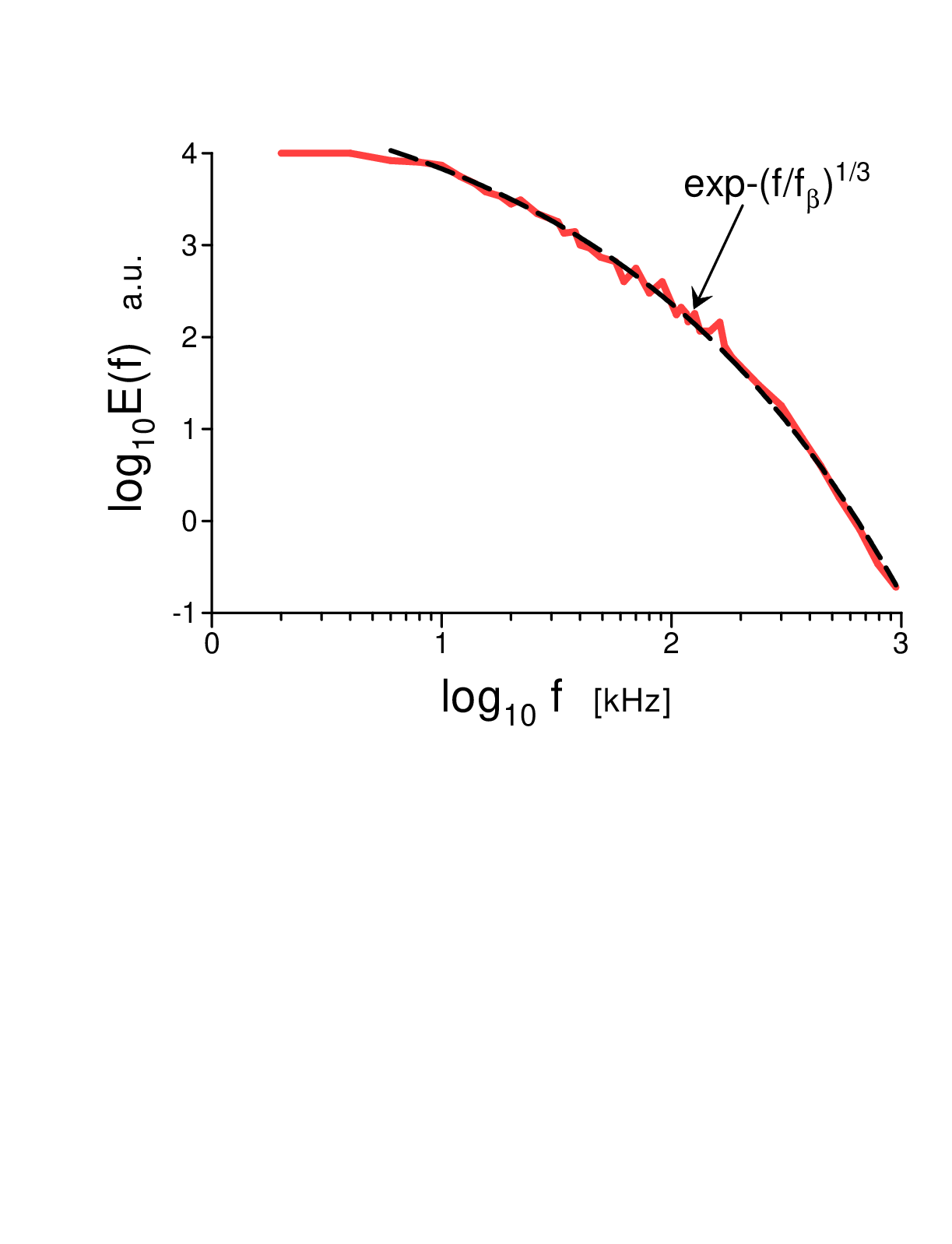} \vspace{-4.6cm}
\caption{As in Fig. 7 but for the  floating potential} 
\end{figure}
%%%%%%%%%%%%%%%%%%%%%%%%%%%%%%%%%%%    
  
   First of all let us estimate the relationship between  $\dot{B}_c$ and $f_c$ similarly to the dimensional estimate Eq. (7)

\begin{equation}
\dot{B}_c \propto \left|\frac{d \rm{ I}_{cr}}{dt}\right|^{1/2} \varepsilon^{-5/4}f_c^{15/4}   
\end{equation}   
i.e., $\alpha = 15/4$ for this case.  Then, for half-normally distributed `Voltage' (i.e. $\dot{B}_c$) from Eq. (14) we obtain $\beta = 15/17$ and the power spectrum

\begin{equation}
 E(f) \propto \exp-(f/f_{\beta})^{15/17}  
\end{equation}  
 for   $\dot{B}(t)$.\\
 
    Figure 7 shows the power spectrum of the time derivative of the toroidal magnetic field measured by internal pick-up coils in the chaotic plasma edge of the RFX-mod reversed field pinch in the Quasi Single Helical state. The spectral data were taken from Fig. 5 of a paper \cite{ago}. The dashed curve corresponds to the power spectrum Eq. (19). \\    
   
    Figure 8 shows the power spectrum of the corresponding floating potential measured by sensors composed of a toroidal array of 72 probes and a poloidal array of 7 probes. The spectral data were taken from Fig. 5 of the paper \cite{ago}.  The dashed curve corresponds to the power spectrum Eq. (17). The consistency between the two results, Fig. 7 and Fig. 8, is clear, as both show the dominance of the second moment of cross-helicity.  \\

\section{Flux of averaged magnetic moment near the separatrix} 

\subsection{Fokker-Planck approach to convective-inertial cascade}

   If collisions are infrequent enough that ions can complete many gyrations in a magnetic field before colliding, the magnetic moment $\mu = m_i v_{\perp}^2 / 2B$ (where $v_{\perp}$ is the ion's velocity perpendicular to the magnetic field, and $m_i$ is the mass of the ion) can be considered an adiabatic invariant, provided that the magnetic field presented as $B$, changes slowly compared to the gyroperiod and over a scale larger than the gyroradius. In the core of the tokamaks and stellarators, this so-called first adiabatic invariant is conserved under the slow-variation assumptions of the guiding-center limit \cite{northrop, kruskal}.\\
   
    While the magnetic moment is an adiabatic invariant in the core, it is not constant in the edge turbulence (near the separatrix) because the separatrix represents a fundamental boundary in phase space where the topology of particle orbits changes (see, for instance, \cite{abd,car,pun,esc} and references therein). As a particle approaches the separatrix, the characteristic frequency of the particle’s motion $\omega_{bounce}$ decreases drastically. Any magnetic field variation, no matter how slow, eventually becomes comparable to the particle’s orbital timescale. Even away from X-points, a shear layer at the separatrix creates a region where "trapped" and "passing" regions of phase space meet. Particles in this layer undergo repeated, small jumps in $\mu$ that accumulate, leading to stochastic (chaotic) motion and the eventual destruction of the invariant. In most fusion plasmas, the separatrix acts as a source of chaos. Particles entering this shear layer will effectively "forget" their initial magnetic moment over a few orbital periods. In a high-shear layer centered at the separatrix, the magnetic field direction changes rapidly over a shear radial distance $L_s$. If the shear is strong enough that $L_s$ approaches the Larmor radius $\rho_L$, the adiabaticity parameter $\rho_L/L_s$, and the particle experiences a ”magnetic shock” within a single gyro-orbit. A combination of the high shear and the separatrix crossing by particles leads to rapid stochastization and particle transport. The breakdown of $\mu$ adiabatic invariance is often not a single event but the result of nonlinear resonance between different periodic motions. Near the separatrix, the particle's bounce motion and its precession motion around the torus begin to happen at frequencies that are integer multiples of each other.  When these resonances "overlap" (according to the Chirikov resonance-overlap criterion), the last remaining KAM (Kolmogorov-Arnold-Moser) surfaces--which act as barriers in phase space--are destroyed. This turns regular, predictable orbits into a global sea of stochasticity. Therefore, in the vicinity of a magnetic separatrix in fusion devices, the magnetic moment $\mu$ ceases to be a strict adiabatic invariant. \\
    
    The Fokker-Planck equation is a standard mathematical tool used to describe how the distribution of these magnetic moments evolves due to small, random perturbations. The kinetic description is reduced via gyro-averaging, a procedure justified by the frequency separation $\omega \ll \omega_{ci}$. By averaging the Fokker-Planck equation over the fast cyclotronic motion, we eliminate the gyrophase dependence and treat the magnetic moment $\mu$ as a formal coordinate. This yields a guiding-center evolution equation where finite Larmor radius effects are implicitly captured within the transport coefficients \cite{hazeltine, helander}. \\
    
     The evolution of the distribution function $f(\bf{x}, \mu, t)$ in the presence of turbulent spatial diffusion is governed by the phase-space Fokker-Planck equation:
\begin{equation}
    \frac{\partial f}{\partial t} + \mathbf{v}_{gc} \cdot \nabla f = \frac{\partial}{\partial \mu} \left( D_{\mu\mu} \frac{\partial f}{\partial \mu} \right) + \nabla \cdot (D_{xx} \nabla f)
\end{equation}
where  $\mathbf{v}_{gc}$ is the gyrocenter velocity vector, $D_{\mu\mu}$ represents the velocity-space diffusion due to non-adiabatic ``kicks'' and $D_{xx}$ represents spatial transport.
    
    This equation needs some additional comments. The experimental observation of broadband turbulence in the frequency range $\omega_{tr} \ll \omega \ll \omega_{ci}$ (where $\omega_{tr}$ is the characteristic frequency of the parallel transit) establishes a specific dynamical regime where the turbulent decorrelation time $\tau_{turb}$ is much shorter than the parallel transit time $\tau_{tr} \approx L_\parallel / v_{th,i}$. The ion thermal velocity is defined as $v_{th,i} = \sqrt{2 k_B T_i / m_i}$, which characterizes the parallel transit frequency $\omega_{tr} \approx v_{th,i} / L_\parallel$ relative to the turbulent fluctuations. In this limit, the ordering of timescales is $\tau_{ci} \ll \tau_{turb} \ll \tau_{tr}$. 

Consequently, the parallel streaming term $v_\parallel \nabla_\parallel f$ in the Fokker-Planck equation is small of order $O(\omega_{tr}/\omega)$ relative to the unsteady and turbulent terms. This justifies a transverse transport model where the ensemble-averaged magnetic moment $\langle \mu \rangle$ evolves as a locally-defined advected-diffusive quantity. In this ``fast-turbulence'' limit, the parallel velocity $v_\parallel$ does not participate in the local radial mixing of the magnetic moment, and the parallel advection term is neglected in the equation because parallel streaming is slow in the near-separatrix region ($L_\parallel \gg v_{th,i}/\omega$), while parallel diffusion and cross-diffusion terms are omitted under the assumption that broadband turbulence primarily drives radial and $\mu$-space transport, effectively decoupling the parallel velocity from the local mixing process in this ``fast-turbulence'' limit (whereas thermalization occurs on a much longer timescale). Additionally, the advection term in $\mu$-space is omitted because the turbulent ``kicks'' are assumed to be stochastic and zero-mean, leading to purely diffusive transport in velocity space without a systematic drift in the magnetic moment.\\

  We define the ensemble average at a spatial location $\mathbf{x}$ as:
\begin{equation}
    \langle \mu \rangle = \frac{\int \mu f (t, {\bf x}, \mu)\, d\mu}{n(t, \bf {x})}, \quad n = \int f (t, {\bf x}, \mu) \, d\mu
\end{equation}
Taking the first moment of Eq. (21) and applying integration by parts (assuming $f (t, {\bf x}, \mu)  \to 0$ at $\mu \to  0$, and $f (t, {\bf x}, \mu)$ decays at $\mu \to  \infty$ faster than any polynomial in $\mu$ -- typically exponentially -- to ensure that total density and energy remain finite), we obtain the equation for the magnetic moment density:
\begin{equation}
    \frac{\partial (n \langle \mu \rangle)}{\partial t} + \nabla \cdot (n \langle \mu \rangle \mathbf{v}_{gc}) = \nabla \cdot (D_{xx} \nabla (n \langle \mu \rangle)) + n \langle \mathcal{A}_\mu \rangle
\end{equation}
where $\langle \mathcal{A}_\mu \rangle = \langle \partial D_{\mu\mu} / \partial \mu \rangle$ is the local source term from non-adiabaticity.

Applying the product rule and subtracting the particle conservation equation ($\partial_t n + \nabla \cdot (n \mathbf{v}_{gc}) = \nabla \cdot (D_{xx} \nabla n)$), we isolate the evolution of $\langle \mu \rangle$:
\begin{equation}
 \hspace{-2cm}   \frac{\partial \langle \mu \rangle}{\partial t}  + (\mathbf{v}_{gc} + \mathbf{v}_{diff}) \cdot \nabla \langle \mu \rangle = \nonumber
 \end{equation}
 \begin{equation}
 \frac{1}{n} \nabla \cdot (n D_{xx} \nabla \langle \mu \rangle)  +  \langle \mathcal{A}_\mu \rangle - \Theta (r-r_{sep}) \frac{\langle \mu \rangle - \mu_{ext}}{\tau_L}
\end{equation}
where $\mathbf{v}_{diff} = -D_{xx} \frac{\nabla n}{n}$ is the diffusive advection velocity and the final term represents the sink -- parallel losses to the divertor in the near-SOL region ($\Theta (r-r_{sep})$ is the Heaviside step function, and $\mu_{ext}$ serves as the target value for the magnetic moment in the external region -- e.g., the Scrape-Off Layer or the wall). The parallel transit time $\tau_L \approx L_{\parallel}/v_{th,i}$. The parallel sink term remains negligible in the outer near-separatrix layer where the connection length $L_\parallel$ is large. In the layer near the separatrix, the magnetic field lines are highly twisted, forcing particles to travel a long distance (typically over 10 meters) along the torus before they reach a physical boundary. This long travel time ensures that the parallel losses are slow compared to the rapid turbulent mixing seen in the measurements. This justifies treating the magnetic moment as a local property that is primarily moved by radial turbulence rather than being immediately lost to the walls.\\

It is important to note that this derivation does not assume an incompressible flow. By defining $\langle \mu \rangle$ as an intensive scalar and utilizing
the full continuity equation, the terms proportional to the plasma compression $\nabla \cdot (\mathbf{v}_{gc} + \mathbf{v}_{diff}$) cancel out identically. Consequently, the resulting transport equation for the transported kinetic moment remains valid for general compressible flows, such as those encountered in the high-gradient region near the separatrix.\\

 Under the assumption of ergodicity and a separation of scales ($\omega_{turb} \ll \omega_{ci}$), we perform a turbulence average $\langle \cdot \rangle_{turb}$ of this equation, removing dependence of the averaged variables on $\bf{x}$ (see below for justification of the ergodicity hypothesis applicability). 

 The evolution of the turbulent-averaged transported kinetic moment $\langle \langle \mu \rangle \rangle_{turb}$ in a turbulent field is governed by the averaged advective term $\left\langle \mathbf{v} \cdot \nabla \langle \mu \rangle \right\rangle_{turb}$, where $\mathbf{v} = \mathbf{v}_{gc} + \mathbf{v}_{diff}$. This term represents the total convective flux of the magnetic moment across the separatrix layer and admits two primary physical interpretations:
The contribution from the mean velocity fields, $\langle \mathbf{v}_{gc} \rangle_{turb}$ and $\langle \mathbf{v}_{diff} \rangle_{turb}$, describes the macroscopic drift of the $\langle \langle \mu \rangle \rangle_{turb}$ profile. This ``background flow'' accounts for the systematic displacement of the kinetic moment by equilibrium ${\bf E} \times {\bf B}$ drifts and the steady-state particle exhaust driven by the density gradient $\nabla n$. In the presence of compressibility ($\nabla \cdot \mathbf{v} \neq 0$), this term ensures that $\langle \mu \rangle$ remains an intensive property, tracking the average value per particle rather than the volumetric density.
The correlation between velocity fluctuations (denoted by the prime symbol) and gradients in the kinetic moment, $\langle {\mathbf v'} \cdot \nabla \mu' \rangle_{turb}$, represents the nonlinear mechanism of turbulent mixing. This term is the ``engine'' of the spectral cascade (variance pumping). It is responsible for ``shredding'' the large-scale spatial inhomogeneities injected by the non-adiabatic source $\langle \mathcal{A}_\mu \rangle$ into increasingly smaller turbulent structures.
   It physically couples the low-frequency plateau (the source region), often observed in the ion saturation current frequency spectrum, to the decaying power range.

Consequently, $\left\langle \mathbf{v} \cdot \nabla \langle \mu \rangle \right\rangle_{turb}$ identifies the effective turbulent convection rate. It is the bridge between the microscopic Hamiltonian ``kicks'' around the separatrix and the macroscopic turbulent dissipation, represented by the turbulent dissipation rate
\begin{equation}
    \varepsilon_\mu \equiv \left\langle \frac{1}{n} \nabla \cdot (n D_{xx} \nabla \langle \mu \rangle) \right\rangle_{turb}
\end{equation}
This term represents the rate at which the turbulent cascade smooths the spatial gradients of $\langle \mu \rangle$, transferring the variance injected by the separatrix source $\langle \mathcal{A}_\mu \rangle$ down to the collisional dissipation scale. 

 The presence of these terms justifies the transition from a deterministic single-particle description to the statistical relaxation of the ensemble, where the stochasticity of the velocity field enforces the observed ergodicity of the $\langle \mu \rangle$ field.\\

     Although $\langle \mu \rangle$ is an intrinsic dynamical variable, its evolution in the turbulent frequency range ($\omega \ll \omega_{ci}$) follows a kind of convective-inertial cascade. The nonlinear coupling between velocity fluctuations and the kinetic moment gradient drives a spectral transfer of $\mu$-variance, where the first-moment dissipation rate $\varepsilon_\mu$ represents the ultimate relaxation of the transported kinetic moment. It is worth noting that while the classical Corrsin-Obukhov approach for {\it passive} scalar turbulent mixing in fluid dynamics focuses on the {\it second}-moment variance cascade to derive spectral power laws, we adopt a {\it first}-moment dissipation rate $\varepsilon_\mu $ derived directly from the Fokker-Planck transport equation. This choice is physically motivated by the following reason. At the transition from 
the variable $v_{\perp}$ to the variable $\mu = m_i v_{\perp}^2 / 2B$ we, already at the kinetic level, introduced a quadratic (energy-like) variable in our description. Then we described the macroscopic relaxation of the magnetic moment profile near the topological singularity of the separatrix, where the mean field and fluctuations are strongly coupled while the  $\langle \mu \rangle$ acts as an {\it active} scalar --  a transported kinetic moment.\\
     
     Numerically, this formulation offers significant advantages: it avoids the high-resolution requirements and sampling noise inherent in tracking second-moment fluctuations, leading to a more computationally robust and stable scheme. By treating $\langle \mu \rangle$ as an active scalar -- a transported kinetic moment strongly coupled to the turbulent field -- we maintain physical consistency with the underlying gyro-kinetics while achieving a more efficient description of the transport dynamics in the high-gradient region of the separatrix.
 Besides, avoiding higher-order moments simplifies the closure problem and reduces the levels of freedom required to describe the relaxation of the plasma state.

\subsection{Ion saturation current dominated by the adiabatic flux of the averaged magnetic moment}

    The numerical fluid simulations, based on the drift-reduced Braginskii equations (and thus assuming strict conservation of the first adiabatic invariant), tend to produce less `intermittent turbulence' than that detected by Langmuir probes. Here we are touching on one of the most debated aspects of edge plasma physics: the transition from drift-wave (electrostatic) dominance to resistive ballooning (electromagnetic) dominance. Near the separatrix, the pressure gradient $\nabla p$ is often at its steepest. In many tokamak scenarios, as we approach the separatrix from the SOL side, the local ratio of plasma pressure to magnetic pressure can approach the threshold for Resistive Ballooning Modes. 
If the drive from the curvature and pressure gradient is sufficiently strong, turbulence develops a strong electromagnetic component. While the authors of the relevant papers often emphasize the measured magnitude of electrostatic flux, they may be underestimating the non-linear coupling where Alfvenic fluctuations "anchor" the turbulence structure, even if the resulting ${\bf E}\times {\bf B}$ transport looks electrostatic on the surface.  If the probe was positioned in a region where the magnetic field lines have high curvature (the "low-field side" or outboard midplane), the fluctuation-induced magnetic flutter can contribute to transport in ways that simple electrostatic probes might not fully capture or might misattribute. In the highly dynamic environment of the separatrix, what is labeled as "electrostatic" density fluctuations in the ion saturation current $I_{sat}$ could actually be a manifestation of electromagnetic/Alfvenic activity (see Introduction). The standard analysis assumes that the spikes in the signals are purely electrostatic density fluctuations. However, the root cause is the electromagnetic displacement of the field line (magnetic flutter). The probe sees the density change, but it is the magnetic field that is "pushing" the plasma onto the probe. There are also induction effects on the probe circuit. The induced currents can contaminate the $I_{sat}$ signal, making it look like a density fluctuation when it is actually an inductive pickup from the electromagnetic turbulence. \\

 Moreover, the quasi-invariance of $\varepsilon_{\mu}$ relies mainly on the electromagnetic turbulence: the particle's "buffeting" near the separatrix is effectively a consequence of their interaction with a turbulent magnetic sea. While electrostatic turbulence is the primary "engine" for moving particles out of the machine, the stochastic evolution of $\mu$ -- the internal "heating" and loss-cone scattering—is the domain of electromagnetic turbulence.\\
 
    While the bulk of the fluctuation energy at low frequencies ($\omega \sim \omega_{bounce}$) may be predominantly electrostatic, the magneto-inertial range is governed by the $\varepsilon_{\mu}$. Because $I_{sat}$ is a parallel flux measurement, it is hypersensitive to the pitch-angle scattering driven by the magnetic "buffeting." Even if the energy density of the electromagnetic fluctuations is orders of magnitude smaller than the electrostatic potential, their superior coupling efficiency to the gyro-phase $\varphi$ allows them to dictate the dynamics of the adiabatic breakdown. Consequently, the observed decay of the $I_{sat}$ power spectrum reflects the constant-rate transfer of the magnetic moment. \\
 
 The observed $I_{sat}$ power spectra, characterized by a flat low-frequency plateau transitioning into a decaying tail, further support this interpretation. The low-frequency constancy represents a statistical stationary, Markovian reservoir of phase-space perturbations that provides a steady-state drive for the ensemble. The subsequent decay then emerges as the dissipative response of the magnetic moment as $\varepsilon_{\mu}$ enforces a constant flux of energy from the gyromotion into the parallel levels of freedom.\\
  
 In this dual-regime framework, the electrostatic turbulence serves as the macroscopic energy reservoir, manifesting as the flat low-frequency plateau of the $I_{sat}$ spectrum. Conversely, the electromagnetic stochasticity acts as the microscopic dissipative sink. While the electrostatic fluctuations dominate the bulk mass transport, the high-frequency decaying tail of the spectrum is dictated by the electromagnetic "buffeting" that breaks the adiabaticity of $\langle \mu \rangle$. Thus, the constant rate $\varepsilon_{\mu}$ represents the steady-state transfer of energy from the electrostatic reservoir into the parallel ion flow, mediated by the superior coupling of magnetic flutter to the particle's gyro-phase.\\
 
  This spectral topology is expected to evolve radially; as the distance from the separatrix (toward the core) increases, the restoration of adiabaticity of $\langle \mu \rangle$ suppresses the electromagnetic sink, causing the electrostatic plateau to become a more profound and extended feature of the spectrum. Conversely, the narrowing of the plateau near the separatrix signals the onset of the stochastic regime, where $\varepsilon_{\mu}$ dictates the rapid transition from ordered electrostatic transport to dissipative parallel loss. For concrete types of tokamaks and stellarators, this phenomenon depends on their design and physical parameters. Another possible mechanism will be considered below.\\
  
  For an illustration, see Figure 9, which shows the power spectra of $I_{sat}$ for some different types of tokamaks and stellarators (cf. Fig. 6 for the corresponding floating potential power spectra). The spectral data were taken from Fig. 3a of the same Ref. \cite{ped}.  The dashed curve in this figure shows a stretched exponential fit for reference. The value of $\beta = 1/2$ will be explained below.\\

 %%%%%%%%%%%%%%% 9 %%%%%%%%%%%%%%%%%%
\begin{figure} \vspace{-0.7cm}\centering
\epsfig{width=.5\textwidth,file=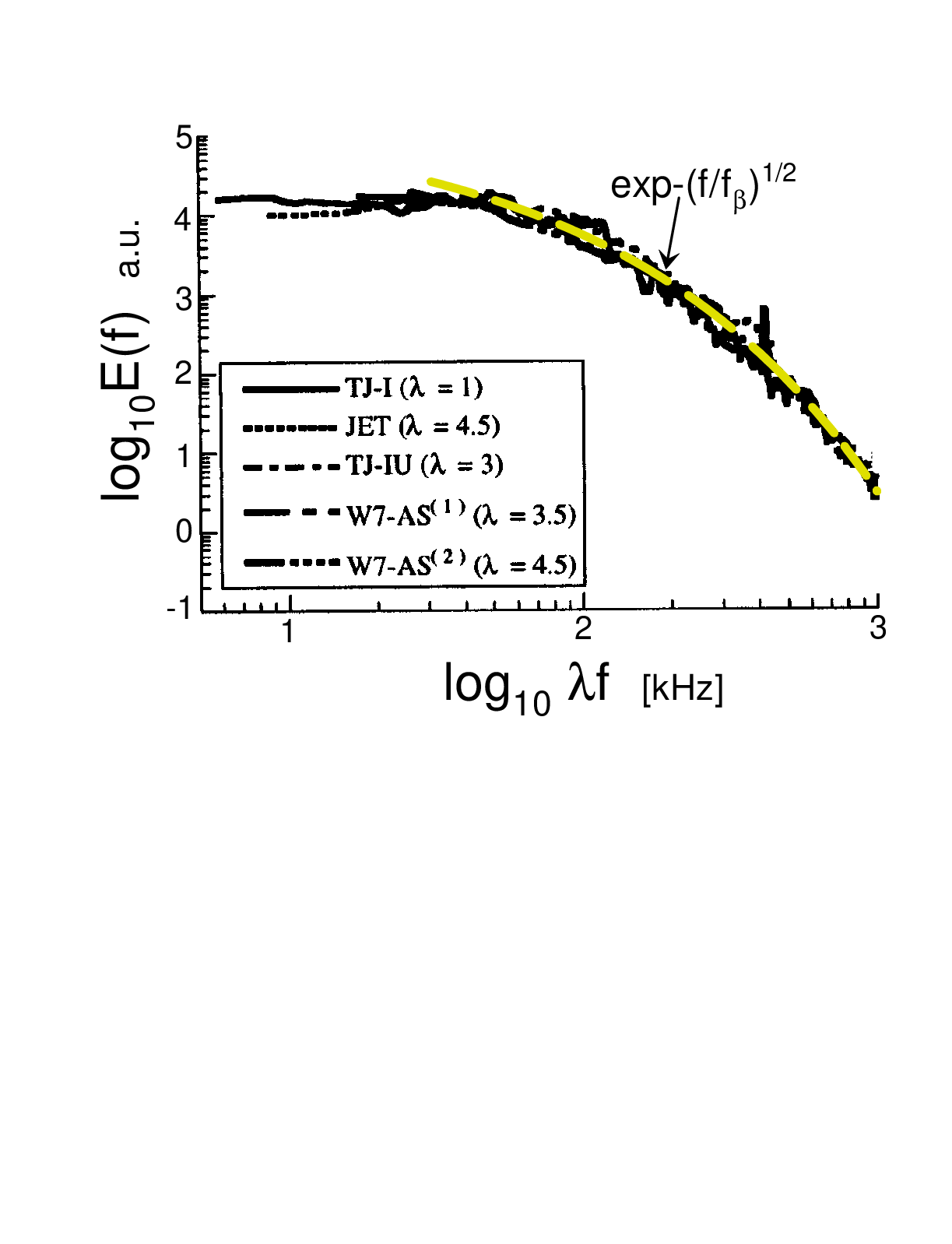} \vspace{-5.6cm}
\caption{Rescaled power spectra of fluctuations of the ion saturation current for different tokamak and stellarator types.} 
\end{figure}
%%%%%%%%%%%%%%%%%%%%%%%%%%%%%%%%%%%  

     Let us summarize the qualitative consequences of the spectra shown in Fig. 9. The presence of a well-defined flat region followed by a stretched exponential decay validates the use of the Fokker-Planck operator. It demonstrates that the transport is neither ballistic nor purely deterministic, but is governed by the statistical relaxation of an {\it ergodic} ensemble in a stochastic magnetic field.
     
     We use $I_{sat}$ to validate the Fokker-Planck approach for $\langle \mu \rangle$, rather than the floating potential. It confirms that $\langle \mu \rangle$ belongs to the scalar/density-like transport family, which is governed by local diffusive laws, rather than the potential/field-like family, which is governed by global electromagnetic constraints. While $I_{sat}$ (as a proxy for density) reflects the local stochastic shuffling of a scalar ($\langle \mu \rangle$) --justifying a Markovian Fokker-Planck treatment -- the potential fluctuations are coupled to long-range magnetic perturbations, masking the local decorrelation scales and highlighting the global nature of the potential field in high-shear regions.\\

  In this context and taking into account previous considerations, the estimation for the characteristic value of the ion saturation current $I_{sat,c}$ can be made using the dimensional considerations
\begin{equation}    
 I_{sat,c} \propto |\varepsilon_{\mu}| \varepsilon^{-1} f_c^2
\end{equation}    
i.e., $\alpha =2$ and it follows from Eq. (14) that $\beta =4/5$. 

  Therefore, the power spectrum of the ion-saturation current $I_{sat}$ is
\begin{equation} 
E(f) \propto \exp-(f/f_{\beta})^{4/5}
\end{equation} 

  This spectrum, although stretched exponential as in Fig. 9, does not provide the observed value $\beta =1/2$. Therefore, we can suppose that a certain additional mechanism is at work here.
  
\section{Strong shear layer and multichannel topologically optimized cascade loop near the separatrix}

\subsection{Constrained variational principle}

   Near the separatrix, the transition from closed to open magnetic field lines creates a localized, high-shear layer where magnetic field topology and plasma gradients interact to generate kinetic helicity through nonlinear interactions. The helicity of a domain with volume $V$ is defined as $H_k = \int_{V} ({\bf u} ({\bf x},t) \cdot  \bm{\omega} ({\bf x},t)) ~ d{\bf x }$, where $\bm{\omega} =\nabla \times {\bf u} $ is the vorticity. In the present case, the volume $V$ should be determined by the radial shear layer width.\\

  Let us construct a hybrid topological quantity 
 \begin{equation} 
  H_{h} = H_k +\lambda_1 H_{cr} + \lambda_2 H_m
 \end{equation}   
 where $\lambda_1$ and $\lambda_2$ are the Lagrangian multipliers which can be used for an optimisation. The magnetic helicities $H_m$ and $H_{cr}$ are ideal MHD invariants, whereas $H_k$ is not. However, in the steady-state system, the topological quantities $H_k$, $H_{cr}$, and $H_m$ are interrelated. Therefore, the self-optimization can make the flux of $H_h$ adiabatic invariant relative to a magneto-inertial range of scales. Namely, the internal leakage of one component, the non-invariant $H_k$, can be exactly compensated by the generation or conversion of others $H_m$ and $H_{cr}$. If $H_k$, $H_{cr}$, and $H_m$ are coupled by the governing dynamics such that their net production/destruction cancels out within the magneto-inertial range, then the hybrid quantity  $H_h$ satisfies the adiabatic flux constancy condition, even if it lacks the formal status of an ideal invariant.
 
 If the shear flow converts kinetic helicity into magnetic helicity at a steady rate across scales, then neither $H_m$ nor $ H_{cr}$ has a constant flux individually. However, the total topological throughput represented by$ H_h$ could remain constant. By finding the right multipliers $\lambda_1$ and $\lambda_2$, we are essentially identifying the eigenmode of the transfer operator. We are defining a coordinate in state-space along which the non-linear "shredding" is balanced by "weaving."
 
   If the non-linear terms redistribute the components of $H_h$ amongst themselves, but preserve the sum $H_h$, the flux 
 becomes a topological constant of the motion for that specific driven system (bypassing non-linearity). This approach essentially describes a steady-state topological flux rather than a classical Kolmogorov-style (scaling) cascade.  The multipliers $\lambda_1$ and $\lambda_2$ essentially define a streamline in the manifold of the system's variables. Along this streamline, the non-linear terms do work, but they do so in a way that preserves the topological value of  $H_h$ as it moves through scale-space (or frequency). We will still preserve the commonly used term ``cascade'' for simplicity.\\
 
   It should be noted that a hybrid helicity (combining $H_m$, $H_{cr}$, and $H_k$) is also considered for the Hall MHD (see, for instance, Ref. \cite{py}). There the hybrid helicity is rigidly optimized to make it an ideal invariant of the Hall MHD.\\ 
 
   To make this type of forward cascade self-optimally stable, the system must have two additional ``slave'' inverse cascades - one of $H_m$ and another of $H_{cr}$. 
   
   The shear pumps $H_k$ and $H_{cr}$, which convert into $H_m$ and move toward smaller scales (forward $ H_h$ master cascade). Small-scale magnetic helicity $H_m$, generated by the non-linear "shredding" of $H_k$, inverse-cascades back toward larger scales. This inverse slave cascade of $H_m$ can back-react on the large-scale shear or the kinetic helicity $H_k$. If the forward master cascade becomes too intense, the resulting surge in large-scale $H _ {m}$ can modify the velocity field (via the Lorentz force) to suppress the original shear-driven production. Usually, stability in turbulence requires dissipation (viscosity/resistivity). In this model, the "slave" inverse cascade acts as a non-dissipative sink for the “master” direct cascade of $H_h$. The small scales don't have to work as hard to dissipate everything because a portion of the topological charge is being recycled back to large scales. This prevents the bottleneck effect (where energy piles up at the dissipation scale), which is a common source of instability in MHD flows. 
   
   By making the inverse cascade a slave of the direct cascade, we have ensured that the return flow of $H_m$ is always proportional to the throughput of the shear layer. If $H_h$ flux increases, inverse flux increases. This creates a self-organizing steady state where the Lagrangian multipliers $\lambda_1$ and  $\lambda_2$ are not just mathematical constructs but represent the physical exchange rate between the two directions of cascade. The system isn't just decaying; it's circulating topological linkage. The stability comes from the fact that the error in one cascade (a fluctuation in flux) is corrected by the response of the other. 
   
   The entire extra $H_m$, which comes to the large-scale reservoir (steady-state), is taken by the forward cascade as a part of $H_h$. That closes the loop. This is a topological recycler. In this scenario, the slave inverse cascade isn't a loss or an accumulation; it’s a feedback supply line. By returning to the large-scale reservoir where it is immediately re-incorporated into the forward flux of $H_h$, the system achieves a robust steady state. Instead of the system constantly needing to drain $H_h$ through small-scale dissipation (which is often turbulent and intermittent), it circulates a portion of the topological charge. This reduces the dissipative load required to maintain stationarity. If the forward cascade of $H_h$ fluctuates upward, it generates more $H_m$ at small scales. This $H_m$ then floods back to the large-scale reservoir, increasing the feedstock for the next cycle of the forward cascade. This acts as a buffer that smooths out the dynamics. Since the inverse cascade is strictly a slave to the forward one, the system can't run away into a pure inverse cascade (which would lead to large-scale instability or condensation). The forward drive of the shear acts as the master regulator. There still exists a small leak of $H_h$ at the very smallest scales to satisfy the second law of thermodynamics.\\
   
   Existence of an additional slave-cascade channel driven by flux $H_{cr}$ completes the topological self-regulating loop. If the $H_k \to H_m$ conversion becomes too violent, the 
$H_{cr}$ flux can act as a governor, smoothing the interaction and preventing the breakdown of the steady state. 

   With $H_{cr}$ slave cascade in the loop, we aren't just recycling "handedness" ($H_k$ and $H_m$) we are recycling the alignment of the fields. The $H_{cr}$ channel ensures that as $H_m$ returns to the large-scale reservoir, it does so in a way that is physically coherent with the existing velocity shear. This prevents the recycling magnetic helicity from clashing with the "old" kinetic flow, which would otherwise trigger instabilities like the Kelvin-Helmholtz or tearing modes.\\
   
   In near SOL the magnetic field effectively stalls the parallel losses (the parallel time $\tau_{\parallel}$ is very large, see above). This creates a topological stagnation zone where the plasma doesn't know it's on open field lines yet. In this region, the model of a closed topological circuit stays valid even outside the separatrix because the radial transport (the cascade) is much faster than the parallel drainage. Though when the feedback from the inverse cascades is slightly weakened (but not yet destroyed by parallel losses), the system tends to organize into solitary topological structures rather than a continuous, stochastic sea.\\
   
  We can conclude that near the separatrix, the steep radial gradients and intense kinetic shear create exactly the environment where the three types of helicity are forced to interact non-linearly. This model of a stabilized, hybrid multichannel topological cascade offers an alternative to standard "turbulence suppression" theories.
  
  If the magnetic helicity $H_m$ generated by small-scale reconnection or kinetic fluctuations inverse-cascades back to the large scales, it doesn't just stay there. It is re-captured by the forward $H_h$ flux. This creates a circulating topological current that can sustain the steep pressure gradient (the Pedestal) without requiring a massive external power increase.
  
  This mutual stabilization explains why the edge can remain in a quasi-steady state despite being driven far from equilibrium. The inverse cascade of $H_m$ acts as a magnetic stabilizer that prevents the kinetic shear from breaking down into purely dissipative, small-scale turbulence. When the $H_{cr}$ flux becomes a steady slave, it greases the interaction between the plasma flow and the magnetic field. This allows the shear to persist without being braked by turbulent Maxwell stresses. It’s not just suppressing turbulence; it's re-aligning it.\\
  
  We can describe the process as a constrained variational principle for a driven-dissipative system: the self-optimization identifies the optimal operating point where the master flux perfectly balances the slave feedback and the dissipative losses, resulting in a statistically stationary state of topological flux (where the Lagrangian multipliers function as the coordinates that define the attractor in the system's phase space).
  
 %%%%%%%%%%%%%%% 10 %%%%%%%%%%%%%%%%%%
\begin{figure} \vspace{-0.8cm}\centering
\epsfig{width=.47\textwidth,file=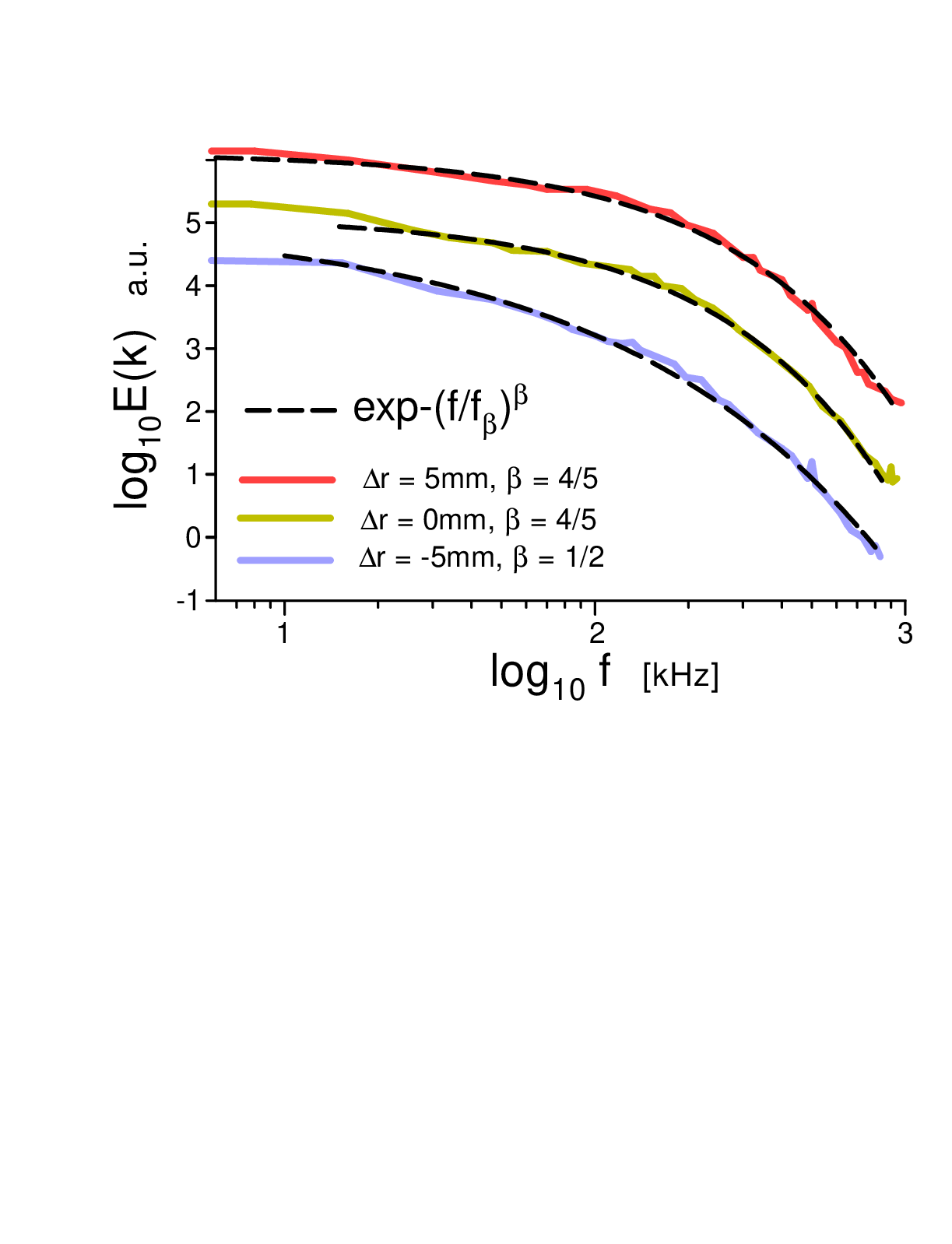} \vspace{-5.1cm}
\caption{Power spectra of the ion saturation current time series measured at three different radial positions (with respect to the separatrix location at $r_{sep}$): $\Delta r =0$  mm and $\Delta r = \pm 5$ mm, where $\Delta r = r-r_{sep}$, for the ISTTOK tokamak, The spectra are vertically shifted for clarity.} 
\end{figure}
%%%%%%%%%%%%%%%%%%%%%%%%%%%%%%%%%%%

\subsection{Frequency power spectra of ion saturation current}

The floating potential $\phi$ is deeply tied to the electric field and, via Ohm’s law, to the magnetic fluctuations and cross-correlations (fluxes of $H_m$ and $H_{cr}$). While the ion saturated $I_{sat}$ is proportional to density (and influenced by the primary velocity fluctuations), it tracks the bulk drive of the system (flux of $H_h$). The different origins of the $\phi$ and $I_{sat}$ determine, in particular, the qualitative difference in their spectral shapes. The low-frequency plateau, typical of the $I_{sat}$ spectrum (Fig. 9), represents the $H_h$ low-frequency reservoir -- here, broadband helicity generation from the shear and magnetic topology is most active and dominant, and can overwhelm any cascade.\\

  For the magneto-inertial range of scales dominated by the flux of $H_h$ (which replaces the energy dissipation rate $\varepsilon$), the estimation Eq. (25) can be replaced by the equation 
\begin{equation}    
 I_{sat,c} \propto |\varepsilon_{\mu}| \left|\frac{dH_h}{dt} \right|^{-1/2} f_c^{1/2}
\end{equation}    
i.e., $\alpha =1/2$ and it follows from Eq. (14) that $\beta =1/2$.  Therefore, the power spectrum of the ion-saturation current $I_{sat}$ in this case is
\begin{equation} 
E(f) \propto \exp-(f/f_{\beta})^{1/2}
\end{equation}    

  Now, one can also look at the appearance of a low-frequency plateau in the frequency spectrum of $I_{sat}$ (preceding the spectral decay Eq. (29)) from another point of view.  The low-frequency reservoir is a place of statistical stationary, broad-band, strong Markovian-like fluctuations, and one can expect that the dynamics in this frequency range should be considered as statistically non-smooth. Therefore, the frequency spectrum in this frequency range should be a power-law (scaling-like, see Introduction). The {\it direct} dimensional considerations for the spectrum, more appropriate for this case, give the equation
\begin{equation}    
  I_{sat,c} \propto |\varepsilon_{\mu}|^2 \left|\frac{dH_h}{dt} \right|^{-1} f^{0}
\end{equation}   
i.e. $\alpha = 0$ and it follows from Eq. (14) that $\beta =0$. That implies a spectral plateau $E(f) \propto constant$ (compare Eqs. (29) and (30) with Fig. 9). \\
           
 For the magneto-inertial range of scales dominated by the flux of the second moment of the hybrid helicity density $d\rm{I_h}/dt$ \cite{mt,ber1,sch}, analogous dimensional considerations result in a scaling relationship 

\begin{equation}
I_{sat,c} \propto |\varepsilon_{\mu}| \left|\frac{d \rm{I_h}}{dt}\right|^{-2/5} \!\!\!\!f_c   
\end{equation}  
i.e., $\alpha =1$ and it follows from Eq. (14) that $\beta =2/3$.  Therefore, the power spectrum of the ion-saturation current $I_{sat}$ in this case is
\begin{equation} 
E(f) \propto \exp-(f/f_{\beta})^{2/3}
\end{equation}  

\section{Measurements of the ion saturation current in fusion devices} 

  Let us consider several examples in addition to those shown in Fig. 9. Figure 10 shows the power spectra of the ion saturation current time series measured at three different radial positions (with respect to the separatrix location at $r_{sep}$): $\Delta r = 0$ and $\Delta r = \pm 5$ mm, where $\Delta r = r-r_{sep}$. The spectral data were taken from Fig. 9 of the Ref. \cite{dud} describing the measurements produced in the ISTTOK tokamak (cf. Fig.1 for the floating potential and the corresponding text).
   The dashed curves in Fig. 10 indicate the best fits corresponding to the stretched exponential Eq. (29) ($\beta = 1/2$) for $\Delta r = -5$ mm, and to Eq. (26) ($\beta =4/5$)  for  $\Delta r =0$ mm and $\Delta r =5$ mm. 
   
 The ion saturation current value of $\beta$ is greater outside the separatrix than inside it; that is, the level of randomness outside is less than that inside.\\
  
 %%%%%%%%%%%%%%% 11%%%%%%%%%%%%%%%%%%
\begin{figure} \vspace{-0.6cm}\centering 
\epsfig{width=.49\textwidth,file=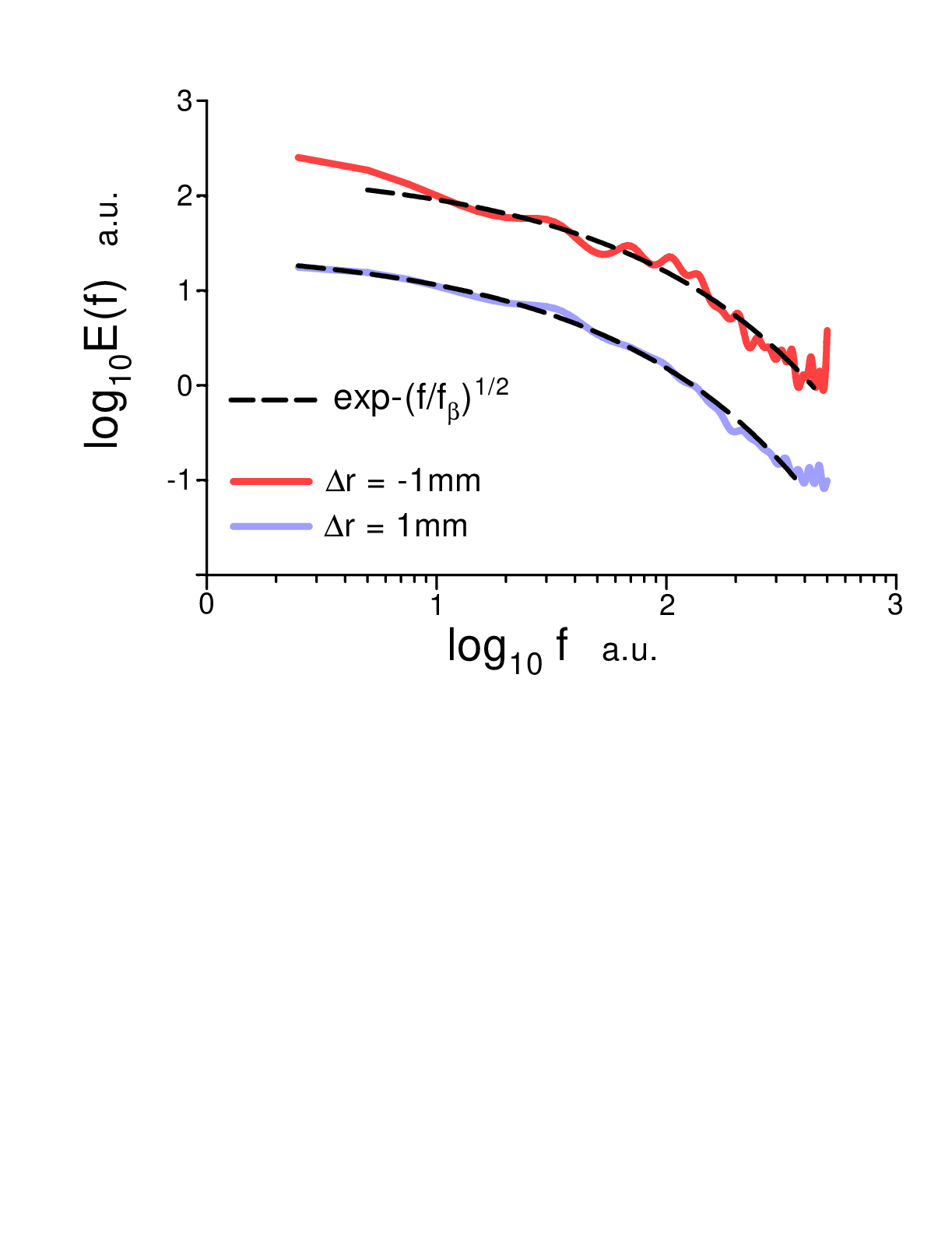} \vspace{-6cm}
\caption{Power spectra of ion saturation current time series measured at two different radial positions (with respect to the separatrix location at $r_{sep}$): $\Delta r = r-r_{sep}$, for W7-AS stellarator.  The spectra are vertically shifted for clarity.} 
\end{figure}
%%%%%%%%%%%%%%%%%%%%%%%%%%%%%%%%%%% 

   The second example comes from the ion saturation current time series measured near the separatrix of the Wendelstein 7-AS (W7-AS) stellarator at H-mode \cite{gri,wag}.
     Figure 11 shows the power spectra of the ion saturation current time series measured at two different radial positions (with respect to the separatrix location at $r_{sep}$): $\Delta r = \pm 1$ mm. The time series were taken from the site of International Stellarator/Heliotron Profile DataBase (ISHPDB)\footnote{\url{https://ishpdb.ipp-hgw.mpg.de/ISHPDB\_public/physicsTopics/edge\_turbulence/index.html}}. The spectra were calculated utilizing the maximum entropy method, yielding optimal resolution for chaotic time series \cite{oh}. 
   
   The dashed curves in Fig. 11 indicate the best fits corresponding to the stretched exponential Eq. (29) ($\beta = 1/2$) for $\Delta r = \pm 1$ mm.\\

   The third example comes from the ion saturation current time series measured near the separatrix of TJ-K stellerator at the confinement magnetic field, $B_0 = 244$ mT (the working gas is Helium) \cite{hor}. Figure 12 shows the power spectra of the ion saturation current time series measured at two different radial positions $\Delta r = -1$ cm and $\Delta r = 2$ cm. The spectral data were taken from Fig. 7 of the paper \cite{hor}. The dashed curves in Fig. 12 indicate the best fits corresponding to the stretched exponential Eq. (29) ($\beta = 1/2$) for both radial positions. \\
   
  %%%%%%%%%%%%%%% 12 %%%%%%%%%%%%%%%%%%
\begin{figure} \vspace{-0.9cm}\centering 
\epsfig{width=.445\textwidth,file=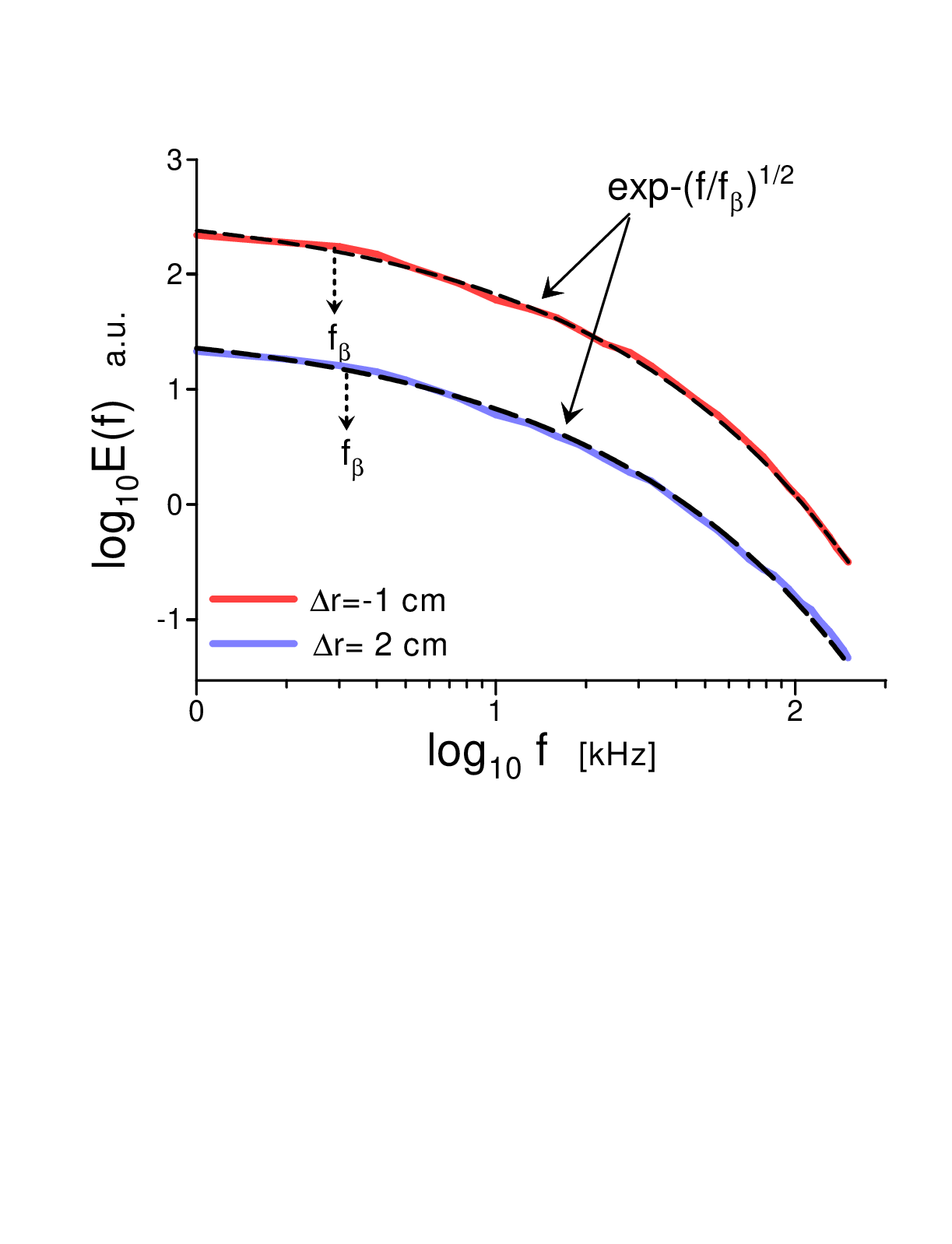} \vspace{-4.2cm}
\caption{Power spectra of ion saturation current measured at two different radial positions (with respect to the separatrix location at $r_{sep}$): $\Delta r = r-r_{sep}$, for TJ-K stellarator.  The spectra are vertically shifted for clarity.} 
\end{figure}
%%%%%%%%%%%%%%%%%%%%%%%%%%%%%%%%%%%
 
 %%%%%%%%%%%%%%% 13 %%%%%%%%%%%%%%%%%%
\begin{figure} \vspace{-0.4cm}\centering 
\epsfig{width=.46\textwidth,file=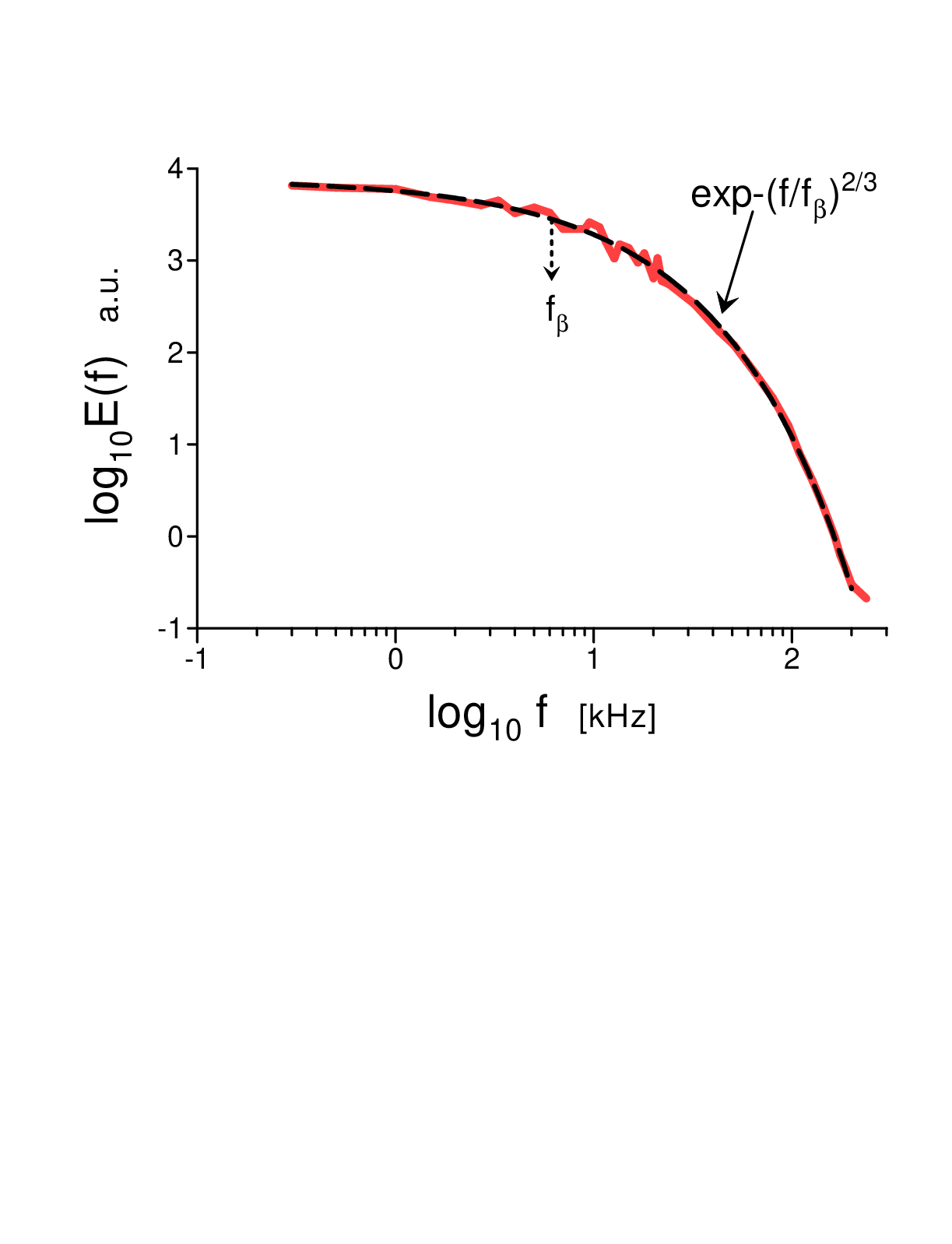} \vspace{-4.6cm}
\caption{Power spectrum of ion saturation current measured outside the separatrix ($\Delta r = 3.7$ cm) for the CDN configuration of MAST tokamak in L-mode.} 
\end{figure}
%%%%%%%%%%%%%%%%%%%%%%%%%%%%%%%%%%%   
   
   Figure 13 shows the Mega Ampere Spherical Tokamak (MAST) power spectra of $I_{sat} (t)$. The time series was obtained at the outboard midplane outside the edge determined by the separatrix. The device operated in L-mode.  The spectral data were taken from Fig. 3a of a paper \cite{hnat}. 
   
 The reference \cite{hnat} examines how magnetic configuration influences the edge turbulence. MAST is frequently operated in a double-null configuration, which features two poloidal field nulls (X-points) close to the last closed flux surface. In a perfect Connected Double Null (CDN) regime, the two separatrices are exactly mapped to one another. This configuration maximizes the interaction with the X-points (the regions of zero poloidal field and infinite shear). In Fig. 13, we can see a steady spectral decay corresponding to the stretched exponential Eq. (32) ($\beta =2/3$) that spans about three decades along the frequency axis (from $f \simeq 0.3$ kHz to $f \simeq 200$ kHz). This occurs due to the CDN's symmetry, which ensures that particles are constantly buffeted by the stochasticity of both the upper and lower X-points. 

\section{Suppression of randomness in the edge turbulence}

\subsection{Physics of suppression}

  In fusion devices like tokamaks and stellarators, the kinetic, magnetic and cross helicities are spontaneously generated or sustained in the plasma shear layers around the separatrix through self-organization (self-optimization) and turbulent processes. The narrow layer centered at the separatrix and populated by helicities works as a low-frequency filter for randomness coming from the core region: it suppresses high-frequency chaos/turbulence, transforming it into low-frequency coherent structures (producing the intermittency), which then propagate into the SOL. From the ``turbulence'' point of view, intermittency is considered a less random state than the fully developed turbulence.\\
  
  Near the separatrix, the transition from closed to open magnetic field lines creates a localized, high-shear layer where magnetic field topology (vortex stretching) and plasma gradients interact to generate kinetic helicity through nonlinear interactions. Kinetic helicity acts as a critical bridge between small-scale turbulence and large-scale self-organization (self-optimization). High kinetic helicity can reduce the nonlinear transfer of energy. This clogs the turbulent cascade, keeping the plasma less random. Also, by feeding energy into the stable Zonal Flows, the helicity effectively deprives the small-scale turbulence of its power. High kinetic helicity is found to reduce turbulent magnetic diffusion, preventing the magnetic topology from becoming tangled and random. The kinetic helicity merges small eddies into large ones, moving energy to coherent scales. Anchors flow to stable skeleton orbits, preventing stochastic drift; it slows magnetic field line mixing, preserving magnetic surface order (diffusion quenching).\\
  
  When reflectional symmetry (parity) is broken in a plasma, the constraints that keep magnetic and cross helicity at zero are lifted, allowing them to be generated spontaneously through turbulent processes. \\
 
 Magnetic helicity can spontaneously emerge when large-scale anisotropy (driven by the strong toroidal field) forces the turbulence into a specific handedness. Once small-scale magnetic helicity is generated (e.g., by micro-instabilities), it tends to undergo an inverse cascade, transferring energy to larger scales. This process is responsible for creating the large-scale coherent magnetic structures that define the edge magnetic topology and reduce the randomness. Magnetic helicity suppresses the {\it randomness} of chaos/turbulence and replaces it with more orderly, self-optimized, helical flows.\\
  
  Near the separatrix, cross helicity is generated primarily through the interaction of gradients. The presence of strong, non-parallel gradients in plasma density and pressure (typical at the edge) acts as a source term for cross helicity. High cross helicity is known to inhibit the nonlinear energy cascade, reducing plasma randomness. Unlike magnetic helicity, which limits the growth of a field, cross-helicity reduces the intensity of the mixing itself: if a velocity fluctuation (an eddy) is perfectly aligned with a magnetic fluctuation, it cannot take hold of the magnetic field to stretch it and create more turbulence. If magnetic helicity is a spatial filter (shredding eddies based on size and shear), cross-helicity is a phase filter: It filters out fluctuations where the velocity and magnetic fields are out of sync. The fluctuations that are phase-locked (Alfv\'{e}nic) have a preference for surviving. This results in the ordered waves we can see near the separatrix, which carry energy without necessarily causing random heat loss. Cross-helicity can actually create non-diffusive transport. While random turbulence always moves heat from hot to cold (diffusion), cross-helicity can drive up-gradient flows or pinch effects. It organizes the plasma to move energy in specific, non-random directions, which can help build the pressure pedestal at the edge.\\
  
  We can consider the near-separatrix layer as a statistical filter. In this view, the self-optimized helical cascades act like a low-pass filter for spatial and temporal randomness. As the highly random turbulence from the core tries to move outward, the intense magnetic and flow shear at the inner boundary shreds those small-scale fluctuations. The self-optimized helical mechanism fastens what remains into more coherent, self-organized, ordered flows. It effectively sifts through the chaotic energy and reorganizes it into the more structured, less random filaments and waves that we eventually see in the outer layer. This filtering is what allows the steep pressure pedestal to form. If the randomness weren't filtered out, the heat would leak out uniformly, preventing the high-confinement states needed for fusion.

 \subsection{Spectral consequences} 
  
  Near the separatrix (both inside and outside) in fusion devices, we have identified four levels of randomeness $\beta = 1, ~4/5,~ 1/2, ~1/3$ for the floating potential ($\phi$) and four levels of randomness for the ion saturation current ($I_{sat}$) $\beta=1,~4/5, ~2/3,~1/2$. The {\it levels} of randomness (characterized by the parameter $\beta$ in Eq. (1)) define not just the bare randomness but also the fundamental physics that governs that randomness. From the previous analysis of the spectral data, the level of randomness inside the separatrix must be greater than or equal to ($\beta$ is less than or equal to) that found outside the separatrix. The {\it level} of randomness $\beta$ characterizes the key parameters determining the process (such as the fluxes) or whether chaos is deterministic ($\beta =1$) or distributed ($\beta <1$). Therefore, we have only four levels of randomness for $\phi$ and four for $I_{sat}$.  For the transition across the separatrix to modify (decrease) the {\it level} of randomness (increase $\beta$), it must change the key parameters that govern the processes (or transform distributed chaos into deterministic). Otherwise, near the separatrix, the {\it levels} of randomness (as we defined them) both inside and outside the separatrix will be equal ($\beta_{inside} = \beta_{outside}$). 
  
  In any way, the layer centered at the separatrix (encompassing the Pedestal, separatrix itself, and the near-SOL) suppresses the randomness coming from the Core and transforms the highly random core's turbulence into intermittent distributed chaos (and even in the deterministic chaos, as in the case shown in Fig. 1).
  
\section{Conclusions and Discussion}

  The topology of the magnetic field is drastically changed in the vicinity of the separatrix in the magnetic fusion devices (such as tokamaks, stellarators, and  RFX-mod reversed field pinch), producing topological chaos near the separatrix. Therefore, in the vicinity of the separatrix, topologically self-optimized fluxes of the shear-generated magnetic and cross helicity (as well as of ensemble-averaged magnetic moment, providing a bridge between kinetics and fluid dynamics represented by shear-generated helicity fluxes) may be central to the plasma's stability, self-organization, and transport properties. The results of the present paper (based on the observed spectra of floating potential, ion saturation current, and magnetic field) support this thesis. This approach bypasses the limitations of simple diffusive models and can be useful for modeling heat-load patterns and particle stochastic ripple loss in next-step fusion reactors.\\
   
   A theoretical framework has been developed for the transport and relaxation of the averaged magnetic moment $\langle \mu \rangle$ near the magnetic separatrix. By treating $\langle \mu \rangle$ as an active scalar, we have accounted for the strong coupling between the kinetic moment profile and the turbulent field in regions of high magnetic shear. The Fokker-Planck transport equation played the pivotal role in closing the gap between kinetic and fluid dynamics descriptions.\\
   
   A synthesis of the theory of the averaged magnetic moment dynamics in turbulent environment with a new multichannel, self-optimized, cascade loop of the helicities generated by a combination of strong shear layer and magnetic topology near the separatrix enables to find spectral laws for the floating potential and ion saturation current, which are in agreement with those measured in numerous experiments with the magnetically confined fusion devises (tokamaks, stellarators, and RFX-mod reversed field pinches).

 The concept of distributed chaos enables a quantitative assessment of the levels of randomness of observed chaotic/turbulent states. The separatrix can separate between the states with different levels of randomness (associated with different dominant helical invariant fluxes). The narrow layer centered at the separatrix (encompassing the Pedestal, the separatrix itself, and the near-SOL, and populated by spontaneously shear-generated helicities of different types) suppresses strong randomness coming from the Core and transforms the highly random Core turbulence into intermittent distributed chaos of the SOL. \\
  
     The edge plasma near the separatrix is not generally dominated by one type of turbulence (electrostatic or electromagnetic), but rather a dynamic interaction where electromagnetic effects, which usually are considered as more destructive to confinement, gain strength as the plasma moves toward higher density and higher performance, potentially becoming the dominant transport mechanism in future reactor devices. That is in the same vein as the advanced EM/Kinetic models used in high-performance fusion research. Therefore, understanding these mechanisms through situations where these effects are locally (near the separatrix) dominant in some currently available devices can be highly advantageous for the future. It is shown that, paradoxically, the floating potential and ion saturation current measured by the Langmuir probes can serve as effective diagnostic indicators of the dominance of electromagnetic turbulence over electrostatic turbulence near the magnetic separatrix.\\
     
     Finally, let us briefly discuss possible applications of the constrained variational principle to active control. 
     
     The control strategy could utilize active helicity injection as a means of prescriptive self-optimization. By deliberately tuning the Lagrangian multipliers through external actuators, we can force the multichannel cascade loop into a specific attractor state. Generally, this state is designed to maximize the stability of the separatrix shear layer, effectively utilizing the topological flux to regulate edge transport and mitigate impulsive dissipative losses. For instance, suppression of edge instabilities (Edge Localized Modes) by injection of cross helicity ($H_c$) and using the self-optimized topological loop as a buffer; controlling the forward cascade of $H_h$ via injected multipliers we might spread the heat more effectively across the scrape-off layer; we could inject the specific helicity ratios needed to lock the plasma into a high-confinement state, instead of waiting for the plasma to do this spontaneously, and so on.\\
     
     Current experimental strategies in fusion devices utilize specific injection techniques that physically implement the constrained variational principle. In the existing experimental contexts, such as Coaxial Helicity Injection (CHI, see, for instance recent Ref. \cite{ots} and references therein) or Local Helicity Injection (LHI, see, for instance recent Refs. \cite{web,scha} and references therein), the Lagrangian multipliers are utilized by internal plasma processes as active control tools. External actuators (e.g., plasma guns or electrodes) prescribe the initial helicity ratios, effectively setting the values of $\lambda_1$ and $\lambda_2$. By tuning these multipliers, the system is forced into a specific attractor state that maintains a stable shear layer, mitigating the Edge Localized Modes and optimizing energy exhaust at the separatrix. 
     
     In high-performance spherical tokamaks, active helicity injection is utilized to reach target attractor states that bypass the need for central solenoidal induction. Let us consider an illustrative example. At Local Helicity Injection (LHI) small, high-current electron sources (plasma guns) are placed in the edge region, injecting current along helical magnetic field lines. This is non-axisymmetric drive that triggers intense magnetic reconnection at the edge. The discrete filaments injected by LHI provide a high-kinetic-helicity ($H_k$) input. The self-optimization process merges these filaments into a slave magnetic structure, identifying the optimal operating point where the individual filamentary linkings relax into a global toroidal flux. High-speed imaging in Pegasus-III and NSTX-U shows that the transition from injected filaments to a smooth plasma current ($I_p$) involves a phase of intense, localized magnetic reconnection. This phase represents the system’s ”calculation” of the Taylor-relaxed state, effectively solving the Euler-Lagrange equations in real-time. In the context of the constrained variational principle, magnetic reconnection acts as the physical mechanism that executes the optimization process. Each reconnection event can be viewed as an optimization step. If the injected fluxes deviate from the optimal hybrid helicity $H_h$, the resulting instabilities trigger reconnection bursts. These bursts transport excess small-scale fluctuations toward the dissipative leak, re-align the velocity and magnetic field vectors, reinforcing the cross-helicity ($H_c$) necessary for dynamic stability, and terminate once the system reaches the variational stationary state, where the drive from the MHD shear is perfectly balanced by the relaxation rate. It serves as the nonlinear solver that drives the system toward the attractor state defined by the Lagrangian multipliers $\lambda_1$ and $\lambda_2$. It moves reconnection from a problem to a solution -- it is the ``tool'' the plasma uses to find the above-mentioned optimal operating point. It explains that when the cascade loop is stable, reconnection activity should reach a steady-state background level (the statistically stationary state). These experiments prove that the plasma is not a passive fluid; it is an active solver. Whether we use CHI, or LHI we are simply providing the boundary conditions for the variational problem, and the plasma uses its internal self-optimized cascade loop to find the most stable solution.\\

{\bf Acknowledgment}

  I thank the HSX and W7-AS teams for sharing their data. I utilized Google Gemini and Claude for grammatical review and to refine the clarity of the text. I reviewed and edited the output and assumes complete responsibility for the content of this publication.\\

\vfill

\end{document}